\documentclass[prb,twocolumn]{revtex4}
              
\usepackage{graphicx}
\usepackage{amsmath}
\usepackage{amsfonts}
\usepackage{multirow}
\usepackage{longtable}
\usepackage{natbib}

\begin{document}

\title{Intertwined Order in a Frustrated 4-leg $t - J$ Cylinder}
\author{John F. Dodaro$^{1,\ast}$, Hong-Chen Jiang$^{2}$, and Steven A. Kivelson$^{1}$}
\affiliation{$^1$Department of Physics, Stanford University, Stanford, CA 94305-4060, USA \\%
$^2$Stanford Institute for Materials and Energy Sciences,
SLAC National Accelerator Laboratory and Stanford University, Menlo Park, California 94025, USA\\%
$^\ast$jdodaro@stanford.edu }
\date{\today}

\begin{abstract}
We report a density-matrix renormalization group study of the $t$-$J$ model with nearest ($t_1$ \& $J_1$) and next-nearest ($t_2$ \& $J_2$) interactions on a 4-leg cylinder with concentration $\delta=1/8$ of doped holes.  We observe an astonishingly complex interplay between uniform $d$-wave superconductivity (SC) and strong spin and charge density wave ordering tendencies (SDW and CDW).  Depending on parameters, the CDWs can be commensurate with period 4 or 8.  By comparing the charge ordering vectors with $2 k_F$, we rule out Fermi surface nesting-induced density wave order in our model.  Magnetic frustration (i.e. $J_2/J_1 \sim 1/2$) significantly quenches SDW correlations with little effect on the CDW.  Typically, the SC order is strongly modulated at the CDW ordering vector and exhibits $d$-wave symmetry around the cylinder.  There is no evidence of a near-degenerate tendency to pair-density wave (PDW) ordering, charge 4e SC, or orbital current order.
\end{abstract}

\maketitle

\section{Introduction}

{\bf Theoretical Context:}
It has become increasingly clear that the phase diagrams of highly correlated electronic systems are inherently complex, with multiple ``intertwined'' ordering tendencies\cite{intertwined} of comparable strength which sometimes appear to ``compete'' and sometimes peacefully coexist.  The defining feature of such systems is that they involve a maximal degree of  quantum frustration; neither the effective kinetic  nor the interaction energy is dominant.  Unfortunately, there are no controlled theoretical methods for solving generic ``intermediate coupling'' problems involving fermions in more than one spacial dimension (1D).  

A partial solution is, however, possible: density matrix renormalization group (DMRG) methods\cite{white_1992, white_1993, white_affleck_scalapino} permit one to efficiently compute the ground-state properties of simple models, such as  the $t$-$J$ and Hubbard models, on ladder systems with the local geometry of any 2D lattice.  Long enough ladders can be treated that confident extrapolation to the thermodynamic limit in one direction is possible, although the method is limited to ladders of at most modest width.  To the extent that the properties we are interested in studying are generic properties of strongly correlated systems, we may hope that the solution of simple models, even where they do not faithfully represent the actual solid state chemistry of any particular material, can teach us something useful.  

With this in mind, we have used the DMRG to explore the behavior of the lightly doped $t$-$J$ model on the 4-leg cylinder with first and second neighbor interactions.  To explore the effects of band-structure on the results, we study a range of values of the ratio of the nearest- and next-nearest neighbor hopping matrix elements, $t_1/t_2$, while the effect of magnetic frustration is explored by varying the ratio of exchange constants, $J_1/J_2$.  Since we are primarily focussing on density wave order, we report results at a fixed density of doped holes, $\delta =1/8$, ({\it i.e.} $1-\delta=7/8$ electrons per site) where these effects are particularly pronounced.

{\bf An Empirical Context:}  Various forms of charge-density wave (CDW) and spin-density wave (SDW) ordering tendencies have been documented in underdoped cuprate high temperature superconductors  (as well as in many other interesting highly correlated materials).\cite{review}  In some cases,\cite{tranquada_1995, otherLSCO, tranquada_2012} as in LSCO, the CDW and SDW orders appear locked to each other in that they occur in the same ranges of ``doped hole'' concentration, $\delta$ (albeit with different onset temperatures) and with mutually commensurate ordering vectors.  In other cases,\cite{ybco_competition, chang_2014, keimer_2014} as in YBCO, the CDW and SDW ordering tendencies seem to compete in that they appear (most strongly) in non-overlapping ranges of $\delta$ and have ordering vectors that are unrelated to one another.  This dichotomy has lead some to conclude\cite{le_tacon_2016} that there are two unrelated ordering tendencies in LSCO and YBCO, despite strong similarities in many other macroscopic signatures of these orders.\cite{JohnHill}  

Basic issues of interpretation have also arisen concerning the driving force that gives rise to these density-wave orders.  From a strong coupling perspective,  CDW formation can arise as a consequence of a mesoscale tendency to phase separation.\cite{zaanen_1989, schulz_1990, machida_1989, physicac, low_1994, zaanen_1998, white_1998, white_1999, oldTroyer, zachar, scalapino_2012}  Density wave order might also arise from specific features of the Fermi surface, including well nested portions of the Fermi surface  (although the validity of this perspective with {\it any} plausible band-structure in more than one dimension has been strongly challenged\cite{mazin_2008}) or with putative ``hot-spots'' on the Fermi surface\cite{hotspottheories_1, hotspottheories_2, hotspottheories_3, hotspottheories_4} spanned by the ordering vector associated with   (probably non-existent) near quantum critical antiferromagnetic fluctuations.

\section{Results}
\begin{figure}
  \includegraphics[width=3.4in]{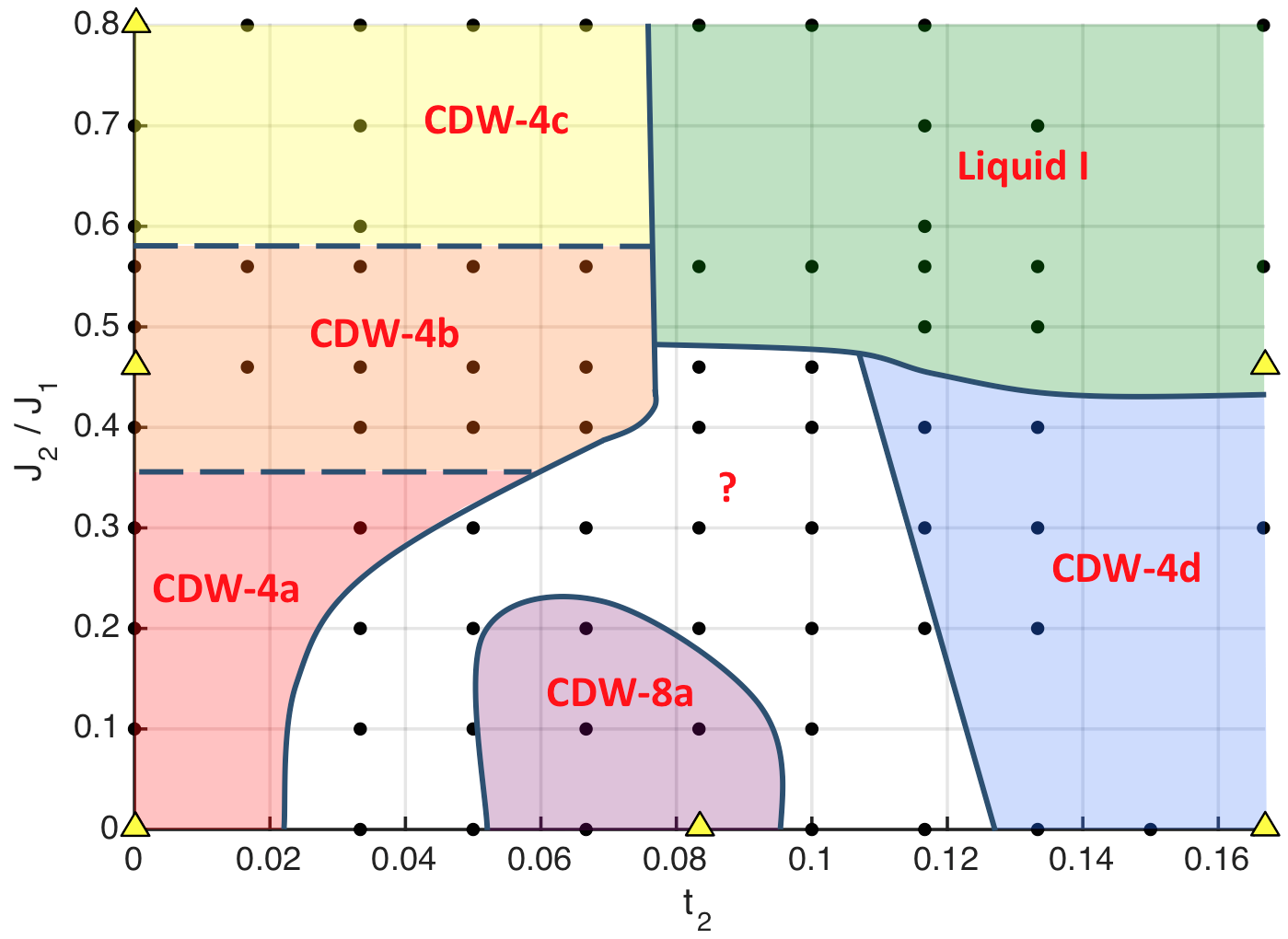}
  \caption{Phase diagram of the frustrated 4-leg $t$-$J$ cylinder with periodic boundary conditions for parameters $t_1 = 1$ and $J_1 = 1/3$.  Solid lines represent schematic phase boundaries based on our data, and dashed lines indicate crossovers.  The nature of the various phases and regimes is represented in Table I, where values for  measured quantities are given for representative points in each region - indicated by the yellow triangles. Values of parameters for which calculations have been carried out are indicated by the dots.}
  \label{PhaseDiagram1}
\end{figure}

\begin{figure}
  \centering
  \includegraphics[width=3.4in]{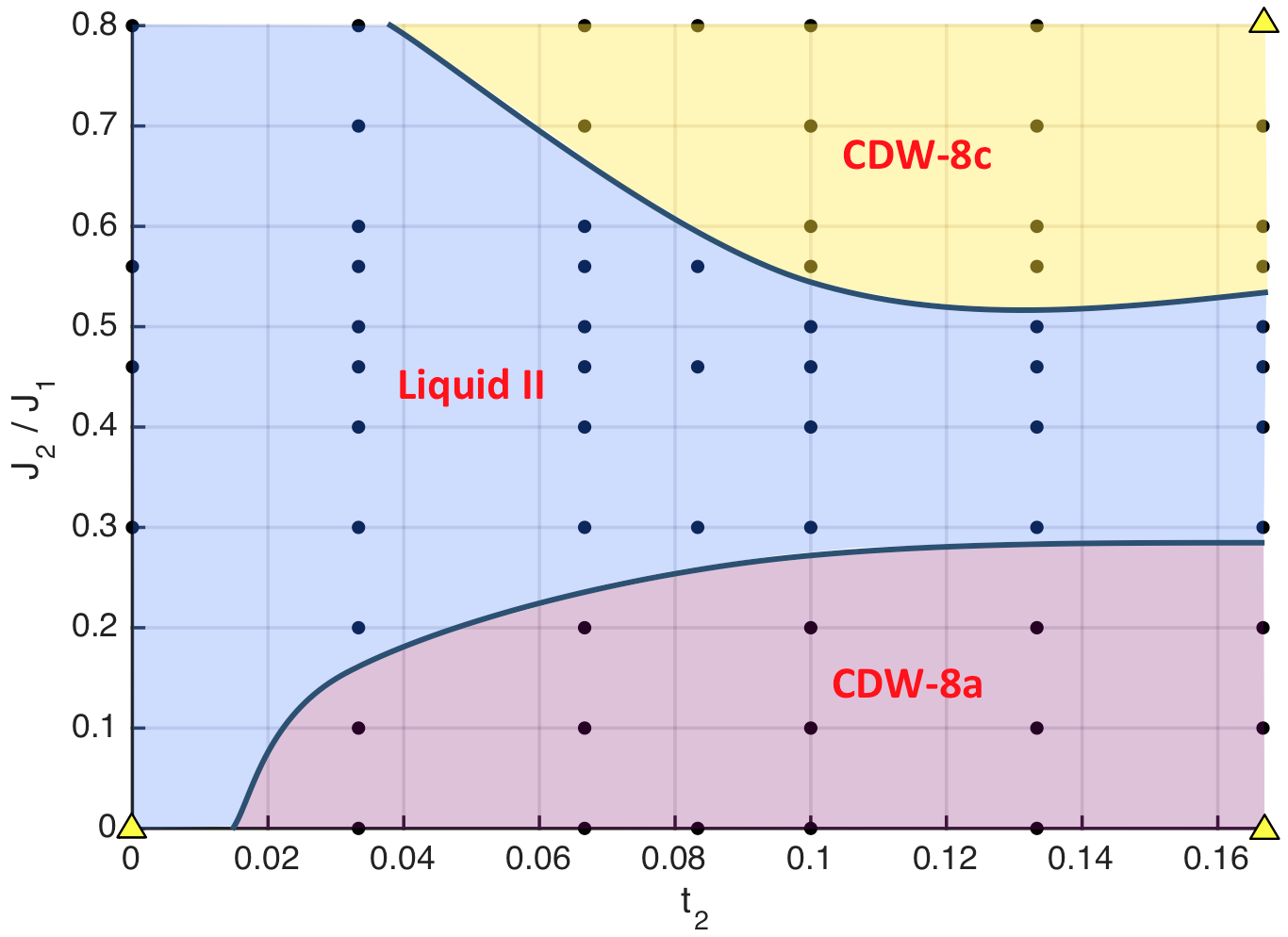}
  \caption{Phase diagram of the frustrated 4-leg $t$-$J$ cylinder with anti-periodic boundary conditions (half quantum of flux) for parameters $t_1 = 1$ and $J_1 = 1/3$.  Symbols are as in Fig. \ref{PhaseDiagram1}.}
  \label{PhaseDiagram2}
\end{figure}

Our principle results are summarized in the schematic phase diagrams in Figs. \ref{PhaseDiagram1} and \ref{PhaseDiagram2} and Table I, which pertain, respectively, to the cases with periodic and anti-periodic boundary conditions around the cylinder.  (Antiperiodic boundary conditions are equivalent to a half-quantum of flux threaded through the cylinder, so these results are referred to as ``flux.'')

{\noindent \bf  1) CDW phases:} We find two distinct patterns of commensurate CDW long-range order: those with period 4 (CDW-4) and with period 8  (CDW-8).  The corresponding ordering vectors are $\vec Q_{cdw}=2\pi(1/4,0)$ for CDW-4 and $\vec Q_{cdw}=2\pi(1/8,0)$ for CDW-8.  Elsewhere in the phase diagram, the CDW correlations fall exponentially, except in the lower parts of the region marked ``?'' where there are weak but possibly power-law incommensurate correlations. 

{\noindent \bf 2) SDW correlations:}  Everywhere, except possibly in Liquid I, the SDW correlations fall exponentially with distance.  However, where the magnetic correlation length, $\xi_{sdw}$, is long compared to $L_y / 2 = 2$ lattice constants, this indicates a strong tendency toward antiferromagnetic order which might well correspond to true long-range-order (LRO) in the 2D limit.  (For comparison, note that the spin correlation length of the 4-leg Heisenberg cylinder\cite{greven} is $\xi_{sdw} = 7a$, a length that increases exponentially\cite{chakravarty} with number of legs of cylinder.)  

We have divided the various phases into regimes according to the nature of the SDW correlations.   
\begin{itemize}
\item {\bf 2a)} In CDW-4a, there is a tendency to SDW order with $\vec Q_{sdw} = 2\pi[3/8,1/2]$;  this is the familiar ``$t$-$J$ stripe'' state discovered by White and Scalapino\cite{white_1998} in the 4-leg ladder with $J_2 = t_2 = 0$ and 8-leg cylinder,\cite{scalapino_2012} which is highly reminiscent of the stripe order discovered  earlier in the 214 family of cuprate high temperature superconductors.\cite{tranquada_1995}  The order here can be visualized as consisting of locally N\'{e}el order with antiphase domain walls coinciding with the location of lines of enhanced doped hole density.  
\item {\bf 2b)} 
With increasing magnetic frustration ($J_2\sim J_1/2$), this phase gives way to CDW-4b, in which the CDW correlations are roughly unchanged, but the magnetic correlations fall exponentially with a length scale equal to a lattice constant or less.  As far as we know, this behavior has not been previously observed in such calculations;  it is more reminiscent of the CDW order seen\cite{ybco_CDW_1, ybco_CDW_2, bscco_CDW} recently in the YBCO and BSCCO families of cuprate high temperature superconductors, where there is no low energy magnetism accompanying the CDW order.  
\item {\bf 2c)} 
For still stronger magnetic frustration ($J_2 \sim J_1$) a new form of short-range ``spin-stripe'' order appears in CDW-4c, with $\vec Q_{sdw} = 2\pi[0,1/2]$.
\item {\bf 2d)}
Increasing to $t_2 \sim t_1 /6$ with small $J_2$ yields CDW-4d, a weak period 4 CDW with short-range $\vec Q_{sdw} = \pi[1,1]$ N\'{e}el antiferromagnetic correlations.
\item {\bf 2e)}  
Similar distinctions exist in CDW-8.  CDW-8a hosts magnetic correlations similar to those found in early Hartree-Fock studies of the Hubbard model\cite{zaanen_1989, schulz_1990};  here $\vec Q_{sdw} = 2\pi[7/16,1/2]$ corresponding to antiphase domain walls in a locally N\'{e}el order.  As far as we know, this behavior has not been previously obtained in any non-mean-field calculations.
\item {\bf 2f)} 
CDW-8c is a period 8 version of CDW-4c, in which  short-range ``spin-stripe'' magnetism appears with $\vec Q_{sdw} = 2\pi[0,1/2]$.
\end{itemize}

{\noindent \bf 3) Superconducting correlations:}  While the Mermin-Wagner theorem precludes superconducting LRO, one might have hoped for power-law superconducting correlations;  this is not seen in any range of parameters we have studied.  However, in most of the phase diagram, short-range superconducting correlations are strong, and the corresponding correlation lengths are longer than $\xi_{sdw}$.  As has been observed in previous studies of the $t$-$J$ model,\cite{white_1997, white_1999, scalapino_2012} the superconducting correlations always have a $d$-wave character in a sense we  make precise in Sec. \ref{pair_symmetry} and Fig. \ref{fig:d-wave}. 

The pair-field correlations additionally exhibit pronounced amplitude oscillations along the cylinder as can be seen, for example, in Fig. \ref{fig:supercor} (b).  These oscillations have the same period as the CDW order, and reflect a strong coupling between CDW and SC order.  We find no evidence of distinct finite momentum pair-density wave (PDW) ordering tendencies, which would be manifest\cite{intertwined} as oscillations with twice the period of the CDW.\cite{pdw_Kondo}  

{\noindent \bf 4) Other correlations:}  Following a suggestion of Affleck\cite{affleck_2007}, we looked for evidence of charge 4e superconducting correlations.  Also, to look for indications of $\vec Q=\vec 0$ orbital loop order\cite{varma_1997} and $\vec Q=\pi(1,1)$ $d$-density wave order,\cite{ddw} we computed the current-current correlation function.  In all cases, these correlation functions fall to below our error limit within a couple of lattice constants.  We thus cannot be more quantitative than to conclude that all of these ordering tendencies appear to be extremely weak in the present model.

{\noindent \bf 5) Relation to ``fermiology:''}  To explore the relation between the CDW ordering vector and the structure of the underlying Fermi surface, we need a prescription for determining the locations of the Fermi points, $\vec k_F$, for any bands that cross the Fermi surface.  ``Bare'' values of $\vec k_F$ can be computed as a function of $t_2/t_1$ by ignoring all interactions, including the $t$-$J$ constraint of no double-occupancy.  The physical (renormalized) locations of Fermi points are identified as values of the crystal momenta at which the fermion occupation probability, $\langle n_{\vec{k} \sigma} \rangle$, has a precipitous drop.  (For example, Fig. \ref{fig:mom_kF} in Supplementary Information.)  In Fig. \ref{fig:2kF}, we compare the values of $2\vec k_F$ computed either way with the observed values of $\vec Q_{cdw}$ and $\vec Q_{sdw}$; we see that {\em there is no obvious relation between the fermiology and the density wave ordering vectors.}

\LTcapwidth=\textwidth  
\setlength{\tabcolsep}{0.47em}
\def\arraystretch{1.67}     
\begin{longtable*}{ | c | c | c | c | c | c | c | c | c | c | }
\hline
& CDW-4a            &  CDW-4b         &  CDW-4c       &   CDW-4d      &  \multicolumn{2}{c |}{CDW-8a}    &  CDW-8c        &  Liq. I             &  Liq. II     \\  \hline
$t_2 / t_1  \ ; \ J_2 / J_1$ 
& 0 ; 0                   &  0 ; 0.46          & 0 ; 0.80         &   1/6 ; 0          &  1/12 ; 0           & 1/6 ; 0 (fl)         & 1/6 ; 0.80 (fl)   & 1/6 ; 0.46       & 0 ; 0 (fl)    \\ \hline 
$\vec{Q}_{cdw}$ 
&  $(1/4,0)$          & $(1/4,0)$          & $(1/4,0)$       &  $[1/4,0]$       &  $(1/8,0)$        & $(1/8,0)$          & $(1/8,0)$        & $-$                 & [1/8,0]      \\ 
$\rho_{cdw}$ $(10^{-3})$         
& $\sim 20$          &$17(1)$             &$10(2)$          & $7(1)$           &$\sim 40$         & $\sim 40$         & $14(3)$          & $0.0(6)$         & $0.0(3)$   \\ 
$\vec Q_{sdw} $ 
& $[3/8,1/2] $       &  $-$                  & $[0,1/2] $       & $[1/2,1/2] $    & $[7/16,1/2] $    & [7/16,1/2]         & [0,1/2]            & $(1/2,0)$        & [5/12,1/2]    \\
$\xi_{sdw}$   																	  
& $3.3(2)$           & $0.80(4)$         & $2.7(1)$         & $3.0(2)$         & $3.7(1)$          & $2.4(7)$          & $\sim 4$         & $6(2)$            & $2.3(5)$   \\  
$\xi_{sc}$      																	 
&  $5.5 $              &  $5.7 $             &  $6.0 $           &  $6.7 $            &  $5.5 $            &  $4.3 $             &  $6.2 $            &  $8.5$           &  $6.6 $  \\ 
$\Phi^{(dw)}_{y,y} / \Phi^{(sc)}_{y,y}$ 
&  0.36                 &  0.40                &  0.48              &  0.34               &  0.40               &  0.49                &  0.97               &  0.22              &  0.31   \\ 
$\Delta_1$ 
&  $0.15$             &  $-$                  &  $-$                &  $0.013$         &  $0.08$           &  $0.183$          &  $-$                & $-$                 &  $0.063$   \\  \hline
\caption{Characterization of the various phases and regimes shown in Figs. \ref{PhaseDiagram1} and \ref{PhaseDiagram2} for $t_1 = 1$, $J_1 = 1/3$, and $L_x = 48$.  For each regime, values of the parameters characterizing all  significant order parameters are given for representative values of $t_2$ and $J_2$;  the notation ``fl'' indicates the case with antiperiodic boundary conditions (flux).  $\vec{Q}_{cdw}$ and $\vec Q_{sdw}$ (measured in units of $2\pi$ with the lattice constant set to unity) correspond to the location of peaks in the density and magnetic structure factors respectively;  wave vectors indicated by $( k_x,k_y)$ correspond to (quasi-)long-range order, while those written as  $[k_x,k_y]$ correspond to short-range order  but with a long enough correlation length that a clearly defined peak in the structure factor can be identified.  $\rho_{cdw}$ is the amplitude of the CDW long-range order extrapolated to the thermodynamic limit ($L_x\to\infty$, see Supplementary Information). $\xi_{sdw}$ is the spin correlation length and $\xi_{sc}$ is the superconducting correlation length. $\Phi^{(dw)}_{y,y} / \Phi^{(sc)}_{y,y}$ is the ratio of the oscillating and uniform components of the pair-field correlation function as in Eq. \ref{eq:phi_PDW}.  $\Delta_1$ is the energy to add one particle to the ladder (see Eq. \ref{Eq:EnergyGap}). } 
\label{tab:phases}
\end{longtable*}

{\bf Road Map:}  In Section \ref{model}, we present the model and define the correlation functions of interest.  In Section \ref{phasediagram}, we present the results from DMRG for the charge and spin correlations in various regions of the ground state phase diagram.  Section \ref{band}  describes calculations of single-particle correlation functions, from which we extract renormalized values of Fermi vectors, $\vec k_F$.  Superconducting correlations and energy gaps are discussed in Sections \ref{superCorr} and Section \ref{energyGap} respectively.  A further discussion of the results is given in Section \ref{discussion}.

\section{The Model}
\label{model}

We treat the $t$-$J$ model Hamiltonian with up to next-nearest neighbor interactions: 
\begin{equation} \label{eq:model}
H = \mathcal{T}_{1} + \mathcal{T}_{2} + \mathcal{H}_{1} + \mathcal{H}_{2}.
\end{equation} 
The kinetic energy term $\mathcal{T}_a$ is:
\begin{equation}
\mathcal{T}_{a} = - t_a \sum_{\sigma,\langle i j \rangle_a } \left(  e^{iA_{ij}}c^\dagger_{i \sigma} c_{j \sigma} + {\rm H.C.} \right) 
\end{equation} 
where $c^\dagger_{i \sigma}$ creates a fermion with spin polarization $\sigma$ at site $\vec{r}_i = (x_i,y_i)$, $A_{ij}$ is the integral of the gauge-field from site $i$ to $j$ (in appropriate units) and 
 $a = 1$ \& $2$ for nearest \& next-nearest neighbor sites respectively. The exchange term $\mathcal{H}_a$ is:
\begin{equation}
\mathcal{H}_{a} = J_a \sum_{\langle i j \rangle_a} \left( \textbf{S}_i \cdot \textbf{S}_j - \frac{1}{4} n_i n_j  \right) 
\end{equation} 
where $n_i = \sum_\sigma n_{i,\sigma} = \sum_\sigma c^\dagger_{i \sigma} c_{i \sigma}$ is the occupation at site $\vec{r}_i$ and $\textbf{S}_i = (1/2) c^\dagger_{i \alpha} \boldsymbol{\sigma}_{\alpha \beta} c_{i \beta}$ for Pauli matrices $\boldsymbol{\sigma}$.  

 There are $N = L_x \times L_y$ sites on the lattice, and the number of electrons would be $N_e = N$ at half filling, {\it i.e.} at $\delta \equiv[N-N_e]/N= 0$.  We will take an electron concentration such that $\delta=1/8$.  The lattice geometry is of 4-leg cylinders ($L_y = 4$) of lengths up to $L_x=48$ with periodic and open boundary conditions along $\widehat{y}$ and $\widehat{x}$ respectively.  In our calculations we set  $t_1 = 1$ as energy unit and $J_1 = 1/3$, and keep up to $m=6561$ number of states in each DMRG block with truncation errors $\epsilon_{\text{tr}} \leq 5 \times 10^{-6}$.
 
When there is no flux through the cylinder, we work in a gauge such that $A_{ij}=0$. For the ``fl'' case, the sum of $A_{ij}$ along any loop that encircles the cylinder once is equal to $\pi$;  so as to work with a real Hamiltonian, in this case we chose a gauge such that $A_{ij}=0$ on all bonds other than those connecting the two sides of one rung of the cylinder, for which $A_{ij}=\pi$.

\underline{${ \langle n_{\vec{k} \sigma} \rangle}$}:  In terms of the Fourier transform of the electron creation operator $c^\dagger_{j \sigma}$ 
\begin{equation}
\begin{split}
c^\dagger_{\vec{k} \sigma} = \frac{1}{\sqrt{N}} \sum_{j} e^{i \vec{k} \cdot \vec{r}_j} c^\dagger_{j \sigma}\ ,
\end{split}
\end{equation}
we can calculate the momentum occupation:
\begin{equation}
\langle n_{\vec{k} \sigma} \rangle = \langle c^\dagger_{\vec{k} \sigma} c_{\vec{k} \sigma} \rangle =  \frac{1}{N} \sum_{i,j} e^{i \vec{k} \cdot (\vec{r}_i - \vec{r}_j)} \big\langle  c^\dagger_{i \sigma} c_{j \sigma}  \big\rangle \ .
\end{equation}

\underline{${ S_{cdw}(\vec{k})}$}:  The CDW structure factor is the Fourier transform of the density-density correlation function
\begin{equation}
\begin{split}
S_{cdw}(\vec{k}) = \frac{1}{N} \sum_{i, j}  e^{i \vec{k} \cdot (\vec{r}_i - \vec{r}_j)  } \big\langle ( n_i - \overline{n} ) ( n_j - \overline{n} ) \big\rangle  \\
\end{split}
\end{equation}
where $\overline{n} = 1 - \delta$ is the average particle density.  Because the ends of the cylinders break translation symmetry, where there is long-range commensurate charge order, its amplitude, $\rho_{cdw}$, can be inferred from the amplitude of the oscillations of $\langle n_i \rangle$ near the middle of the cylinder (see Fig. \ref{fig:amp_fit} in Supplementary Information for detail), extrapolated to the $L_x \to \infty$ limit;  the wave-vector of these oscillations, $\vec Q_{cdw}$, always corresponds to a pronounced peak in the structure factor $S_{cdw}(\vec{k})$.  
 
 \underline{ ${ S_{sdw}(\vec{k})}$}:  The SDW structure factor is
 \begin{eqnarray*}
&&S_{sdw}(\vec{k}) = \frac{1}{N} \sum_{i, j} e^{i \vec{k} \cdot ( \vec{r}_i - \vec{r}_j ) }  \big\langle \textbf{S}_i \cdot \textbf{S}_j  \big\rangle  \ .
\end{eqnarray*}
We obtain the correlation length, $\xi_{sdw}$, by fitting the long-distance fall-off of spin-spin correlation function, $\big\langle \textbf{S}_i \cdot \textbf{S}_j  \big\rangle$, to an exponential.  This approach fails only if the correlation length is comparable or longer than  the system size, or if the correlations fall with a power law.\cite{xi_SDW}  We use the somewhat arbitrary criterion, $\xi_{sdw} > L_y / 2 = 2$, as the definition of ``significant'' SDW correlations.

\underline{$ \Phi_{sc}$}:  The singlet superconducting pair-field creation operator on neighboring pairs of sites is defined as
\begin{equation}
\phi_{\vec{a}}^\dagger(\vec r_i) = \frac{1}{\sqrt{2}} \left( c^\dagger_{i \uparrow} c^\dagger_{i + a \downarrow} - c^\dagger_{i \downarrow} c^\dagger_{i + a \uparrow} \right)
\end{equation}
where $\vec{a} = \widehat{x}$, $\widehat{y}$, and $\widehat{x} + \widehat{y}$ corresponds to different singlet orientations.  The primary diagnostic of SC order we have analyzed is the pair-field correlator,
\begin{equation} \label{eq:SCPairField}
\Phi_{a,a^\prime}(\vec r,\vec r^{\ \prime}) = \big\langle \phi_{\vec{a}}^\dagger(\vec r) \phi_{\vec{a}^{ \prime} }(\vec r^{\ \prime})   \big\rangle 
\end{equation}
which in all cases we have studied, for large separations along the ladder,   $1 \ll |x| \ll L_x$ where $ \vec r-\vec r^{\ \prime}  = x\hat x\equiv \vec X$, $\Phi$ can be expressed as the sum of a uniform and an oscillatory piece as
\begin{equation} \label{eq:phi_PDW}
\Phi_{a,a^{\prime}}(\vec r,\vec r^{ \ \prime}) \sim \big\{\Phi^{(sc)}_{a,a^{\prime}} + \Phi^{(dw)}_{a,a^{\prime}} \cos[\vec Q_{cdw}\cdot\vec X+\theta_0]\big\} e^{-|\vec X|/\xi_{sc}},
\end{equation}
from which we derive the values for $\Phi^{(sc)}_{y,y} / \Phi^{(dw)}_{y,y}$ and $\xi_{sc}$ quoted in Table I and elsewhere.  If there are superconducting correlations that oscillate with any other wave-vectors, {\it i.e.} any PDW ordering tendencies, they are either too weak or too short-ranged to be observable in our calculations.  We have looked for SC correlations in a complimentary fashion by applying an external field $D_{\vec{a}}(\vec{r}_i) \sim \left( \phi^\dagger_{\vec{a}}(\vec{r}_i) + \phi_{\vec{a}}(\vec{r}_i) \right)$ along one rung of the ladder, and then measured the induced pair-field expectation values $\langle D_{\vec{a}}(\vec{r}_i) \rangle$, as a function of distance from this rung. Results obtained in this manner are broadly consistent with those obtained from the analysis of $\Phi$.

\underline{Other Orders}: 
To search for charge $4e$ superconducting correlations, we have computed the correlation functions of the two singlet ``quartet'' creation operators
\begin{equation} \label{eq:SC4eEVEN}
\begin{split}
\phi^\dagger_{4e,+} 
 = \frac{1}{2} \Big( c^\dagger_{1 \uparrow} c^\dagger_{2 \downarrow} c^\dagger_{3 \downarrow} c^\dagger_{4 \uparrow} - c^\dagger_{1 \downarrow} c^\dagger_{2 \uparrow} c^\dagger_{3 \uparrow} c^\dagger_{4 \downarrow} \\
 - c^\dagger_{1 \uparrow} c^\dagger_{2 \uparrow} c^\dagger_{3 \downarrow} c^\dagger_{4 \downarrow} - c^\dagger_{1 \downarrow} c^\dagger_{2 \downarrow} c^\dagger_{3 \uparrow} c^\dagger_{4 \uparrow} \Big) 
\end{split}
\end{equation}
\begin{equation} \label{eq:SC4eODD}
\begin{split}
\phi^\dagger_{4e,-} 
 = \frac{1}{\sqrt{12}} \Big( 2 c^\dagger_{1 \uparrow} c^\dagger_{2 \downarrow} c^\dagger_{3 \uparrow} c^\dagger_{4 \downarrow} + 2 c^\dagger_{1 \downarrow} c^\dagger_{2 \uparrow} c^\dagger_{3 \downarrow} c^\dagger_{4 \uparrow}    \\
 - c^\dagger_{1 \uparrow} c^\dagger_{2 \downarrow} c^\dagger_{3 \downarrow} c^\dagger_{4 \uparrow}  - c^\dagger_{1 \downarrow} c^\dagger_{2 \uparrow} c^\dagger_{3 \uparrow} c^\dagger_{4 \downarrow}  \\
  - c^\dagger_{1 \uparrow} c^\dagger_{2 \uparrow} c^\dagger_{3 \downarrow} c^\dagger_{4 \downarrow}   - c^\dagger_{1 \downarrow} c^\dagger_{2 \downarrow} c^\dagger_{3 \uparrow} c^\dagger_{4 \uparrow} \Big) 
\end{split}
\end{equation}
which are defined on each plaquette around which we label the sites  1-2-3-4 starting in the bottom left corner and moving counter-clockwise.

To check for possible loop current or $d$-density wave order, we measured the current-current correlation function for current operator $J^\sigma_{pq} = - i ( c^\dagger_{p, \sigma} c_{q, \sigma} - c^\dagger_{q, \sigma} c_{p, \sigma}  )$ with sites $p$ and $q$ along $\widehat{x}$ and $\widehat{y}$ bonds and found exponentially decaying correlations with $\xi_{loop} < 2$ across the phase diagram (see Fig. \ref{fig:loop} in Supplementary Information) indicating loop current order is unlikely in our model within the range of parameters explored.

\section{Phase Diagram}
\label{phasediagram}

In the limit $\delta =0$, the present model reduces to an insulating spin-1/2 Heisenberg quantum antiferromagnet, so it is reasonable to view the present system as being a doped antiferromagnet.  The ground-state of the spin 1/2 Heisenberg antiferromagnet on the 2D square lattice with $J_2/J_1$ small is known to have Neel ($\vec Q_{sdw}=[\pi,\pi]$) order, while for large enough $J_2/J_1$ it has ``stripe antiferromagnetic'' ($\vec Q_{sdw} = [\pi,0]$ or $[0,\pi]$) order. For $J_2/J_1 \sim 1/2$, previous studies\cite{figuerido_1989, jiang_2012, donnasheng, wen_2016} have established the existence of a quantum paramagnetic phase with a  spin-gap and tentatively concluded\cite{jiang_2012} that there is a spin liquid for $J_2 / J_1 = 0.41 - 0.5$ and a valence bond solid for $J_2 / J_1 = 0.5 - 0.62$.  Thus, the doped ladder with $J_2 / J_1 \sim 1/2$  can be viewed as a doped quantum paramagnet.
 

The ground-state phase diagrams for $\delta=1/8$ extracted from our DMRG results are shown in Fig. \ref{PhaseDiagram1} and \ref{PhaseDiagram2}, and reported in Table \ref{tab:phases}.  In this section we give a taste of the explicit analysis that leads to the conclusions summarized there.

\begin{figure}
  \centering
  \includegraphics[width=3.4in]{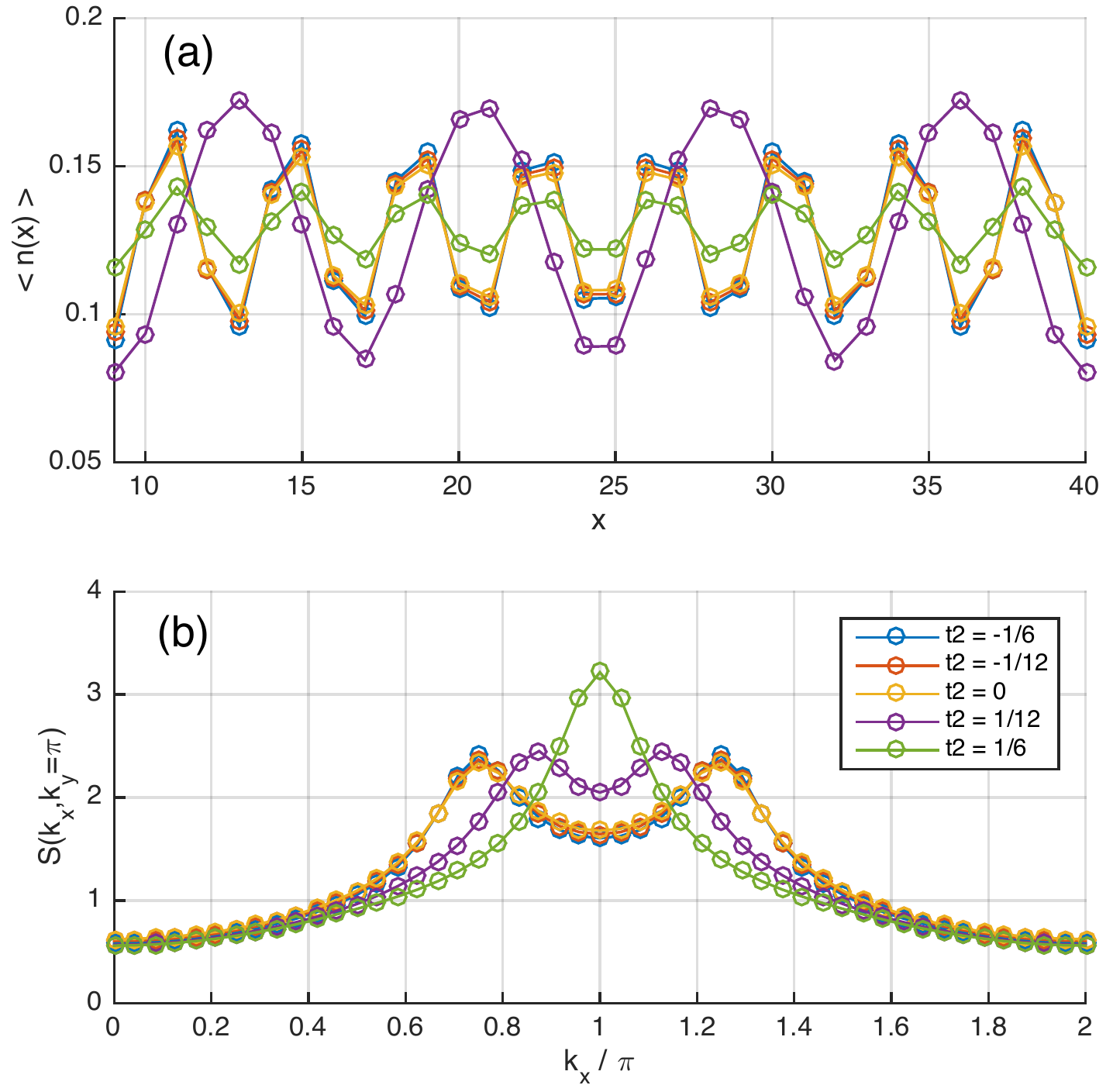}
  \caption{(a) Density profile and (b) spin structure factor at $k_y = \pi$ for $J_2 = 0$ and different $t_2$, for the $t$-$J$ model without flux.}
  \label{fig:den_spin_j2-0_L48}
\end{figure}

{\underline{\bf Analysis for $J_2\approx 0$:}}
The density profile $\langle n(x) \rangle = (1 / L_y) \sum_y \langle n(x,y) \rangle$ for $J_2 = 0$ and different values of $t_2$ are shown in Fig. \ref{fig:den_spin_j2-0_L48} (a) for the central 32 sites of the $48 \times 4$ system.  The commensurate period 4 CDW-4a with ordering vector $\vec Q_{cdw}  = 2\pi(2 \delta , 0)$ is stable at least for $-1/6 \leq t_2  \lesssim 0.02$.  Expressed in terms of a picture of charge-stripes, there are 2 holes per stripe or $\rho = 1/2$ holes per domain wall unit cell (``half-filled stripes'').  The largest CDW amplitude occurs for the period 8 ($\vec Q_{cdw} = 2\pi( \delta , 0)$, {\it i.e.} $\rho=1$) in a range of $t_2 \sim 1/12$ identified as  CDW-8a.  For larger $t_2$, the period 4 CDW order is weakened; we refer to this phase as CDW-4d.

The $J_2 = 0$ spin structure factor as a function of $k_x$ is shown for different $t_2$ in Fig. \ref{fig:den_spin_j2-0_L48} (b) for $k_y = \pi$.  In CDW-4a  (for small $t_2 > 0$) the magnetic structure factor has a pronounced peak $\vec Q_{sdw} = \vec Q_{AF} \pm \Delta \vec Q$ where $Q_{AF}  = \pi(1,1)$ and the antiferromagnetic incommensurability is $\Delta \vec Q  = \pi (2 \delta, 0)$.  CDW-8a ($t_2 \sim 1/12$) has local magnetic correlations corresponding to a SDW with antiferromagnetic incommensurability $\Delta \vec Q = \pi(\delta, 0)$.  The antiferromagnetic incommensurability disappears for $t_2 = 1/2$ (CDW-4d) where the system exhibits short-ranged N\'{e}el ordered with spin-spin correlation length $\xi_{sdw} \sim 3.0$.

Turning now to the cylinder threaded with a half-flux quantum ({\it i.e.} the results in Fig. \ref{PhaseDiagram2}) for $t_2 \approx 0$, charge order vanishes in the $L_x \rightarrow \infty$ limit. The magnetic structure factor exhibits a peak away from $\vec{Q}_{AF}$ with antiferromagnetic incommensurability $\Delta \vec Q = \pi(1/6, 0)$;  as these correlations are exponentially decaying, we identify this phase as Liquid II to draw a distinction with Liquid I which may possess power-law correlations.  Upon increasing $t_2$, the CDW-8a phase is stabilized with a large amplitude that increases with $t_2$.

\begin{figure}
  \centering
  \includegraphics[width=3.4in]{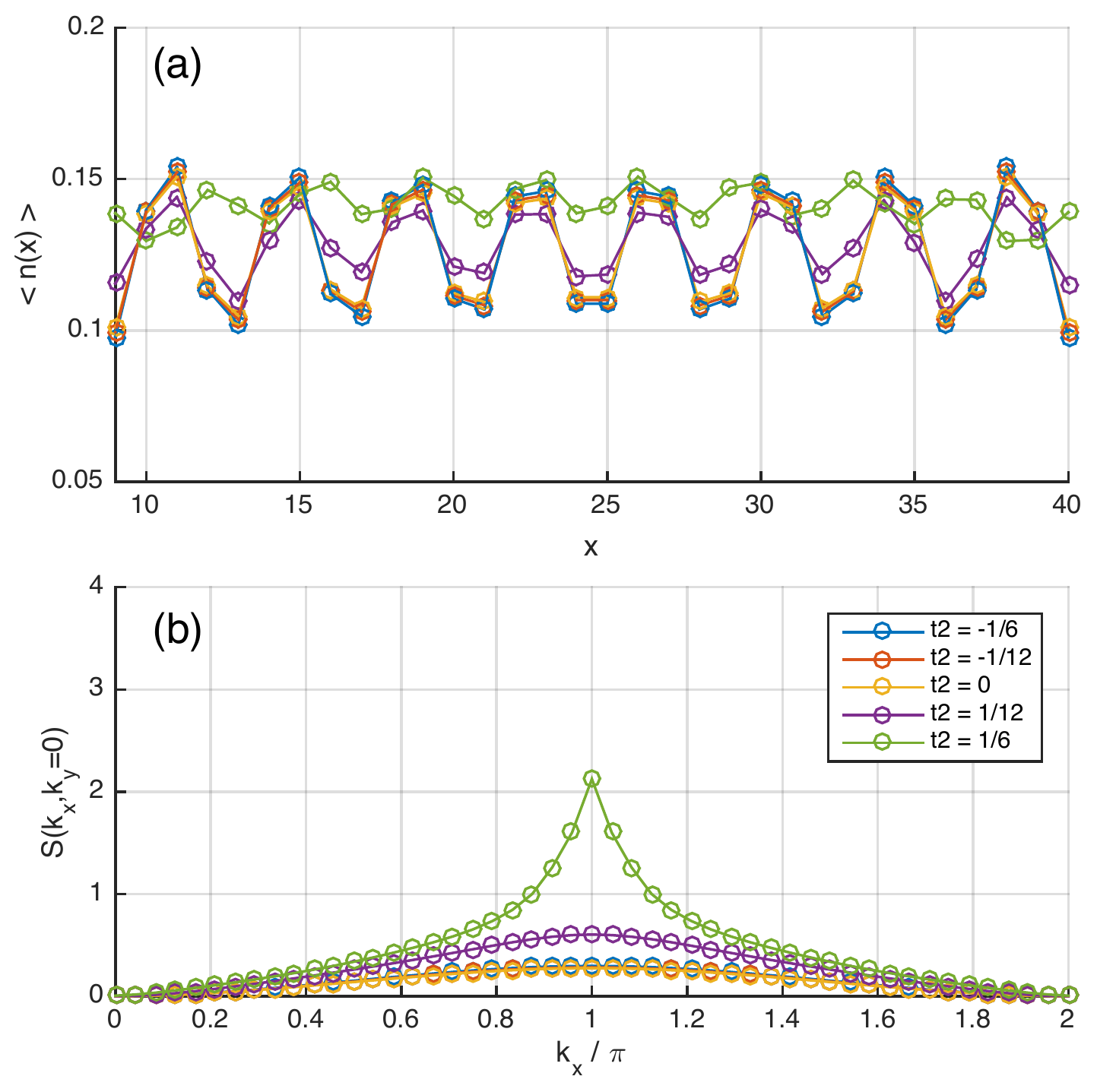}
  \caption{(a) Density profile and (b) spin structure factor at $k_y = 0$ for $J_2 = 0.46 J_1$ and different $t_2$, for the $t$-$J$ model without flux.}
  \label{fig:den_spin_j2-46_L48}
\end{figure}

{\underline{\bf Analysis for $J_2\sim J_1/2$:}}
Next, we treat $J_2 = 0.46 J_1$, corresponding to the quantum paramagnet phase in the undoped 2D system. We find a nearly identical period 4 CDW for $-1/6 \leq t_2 \lesssim 1/12$ in Fig. \ref{fig:den_spin_j2-46_L48} (a), but a vanishing of SDW order as seen in Fig. \ref{fig:den_spin_j2-46_L48} (b) with $\xi_{sdw} < 1$;  this corresponds to  CDW-4b in Fig. \ref{PhaseDiagram1}.  The period 8 CDW is absent for $J_2 = 0.46 J_1$ at $t_2 \sim 1/12$, and for $1/12 < t_2 \leq 1/6$ we find quasi-long-ranged $(\pi,0)$ anti-ferromagnetism and vanishing CDW order; we refer to this phase as Liquid I.  Similar features were seen for $J_2 = 0.56 J_1$ where in the undoped 2D case there may be a valence bond solid state.     

Turning again to the flux case, we find the Liquid II phase for $0 \leq t_2 \leq 1/6$ which lacks pronounced peaks in both density and magnetic structure factors.

\begin{figure}
  \centering
  \includegraphics[width=3.4in]{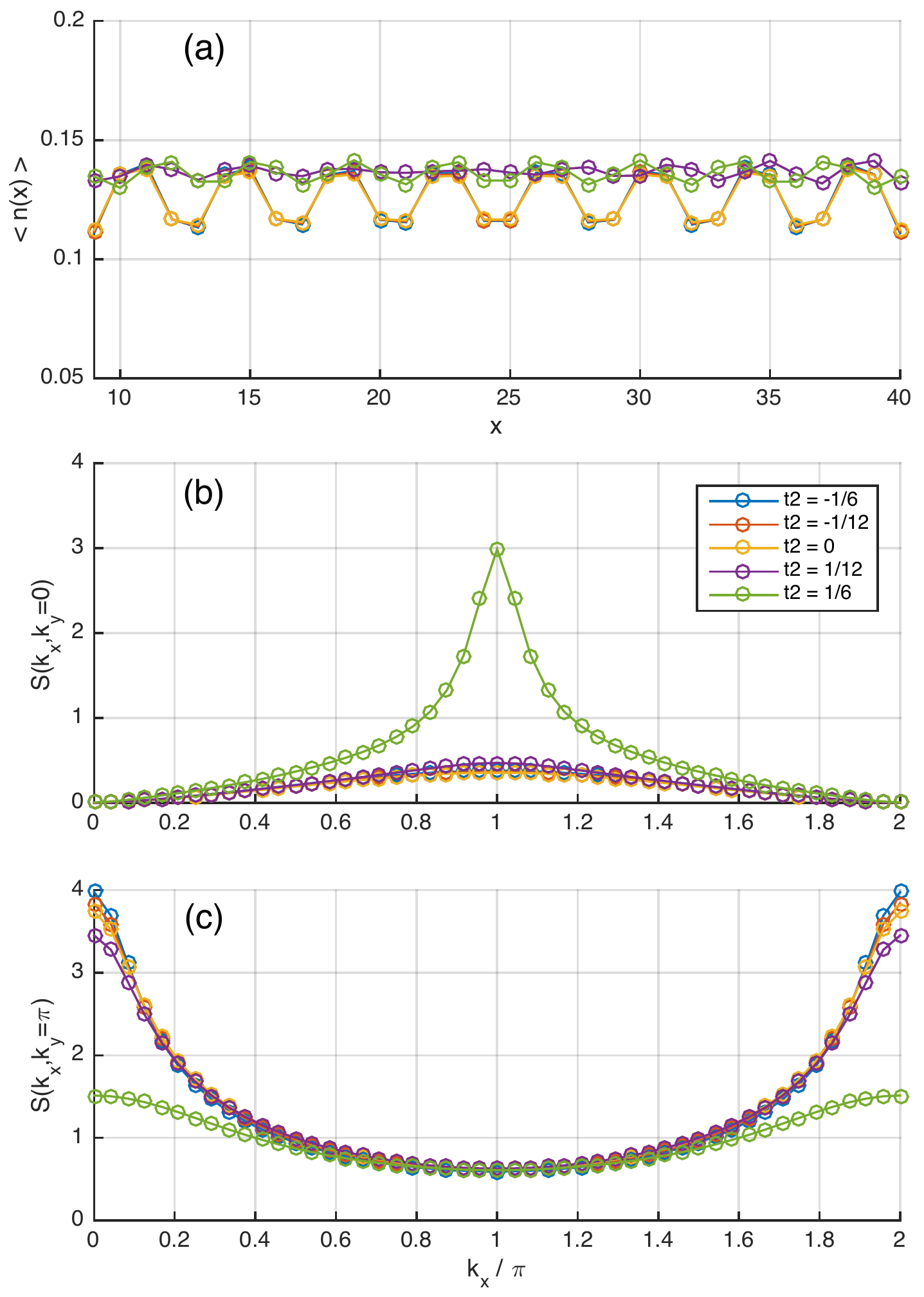}
  \caption{(a) Density profile and spin structure factor at (b) $k_y = 0$ and (c) $k_y = \pi$ for $J_2 = 0.80 J_1$ and different $t_2$, for the $t$-$J$ model without flux.}
  \label{fig:den_spin_j2-08_L48}
\end{figure}

{\underline{\bf Analysis for $J_2 \sim J_1$:}} 
In the 2D undoped system, $J_2 = 0.80 J_1$ lies in the phase exhibiting stripe anti-ferromagnetism.  Here we find a weaker period 4 CDW for $-1/6 \leq t_2 \lesssim 1/12$ and a short-ranged $[ 0,\pi ]$ SDW (labelled as CDW-4c in Table \ref{tab:phases}).  Increasing $t_2 > 1/12$ results in quasi-long-ranged $( \pi,0 )$ SDW which again shows no apparent CDW order (Liquid I phase) as seen in Fig. \ref{fig:den_spin_j2-08_L48}. 

Turning to the flux case, we find a period 8 CDW for $t_2 \gtrsim 1/30$ accompanied by a short-ranged $[ 0,\pi ]$ SDW labelled as CDW-8c phase.

\section{Band Structure}
\label{band}

\begin{figure}
  \centering
  \includegraphics[width=3.2in]{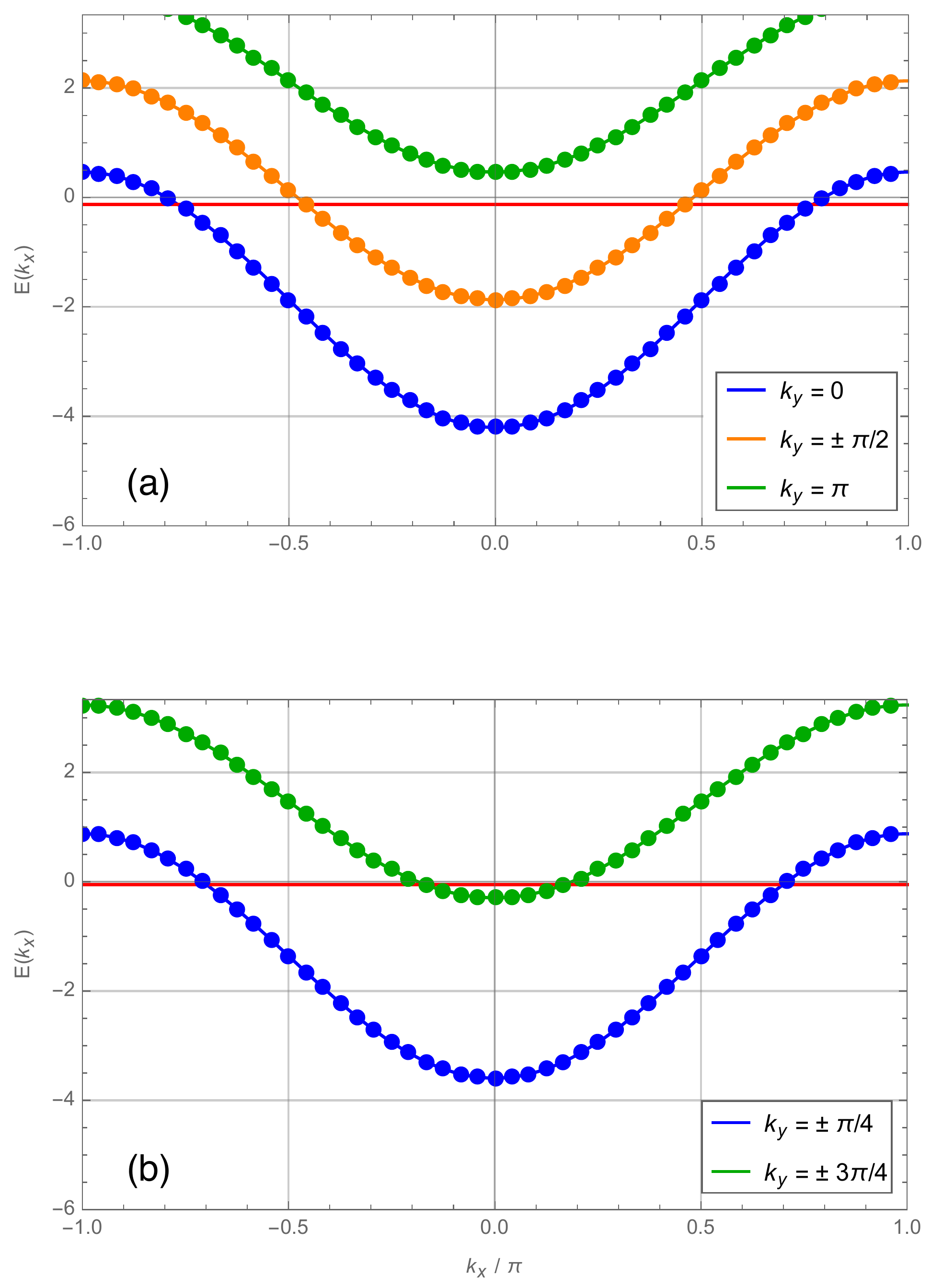}
  \caption{Band structure of non-interacting model (see Eq. \ref{eq:bandstructure}) at $t_1 = 1$, $t_2 = 1/12$, $\delta = 1/8$ with (a) periodic and (b) antiperiodic boundary conditions.}
  \label{fig:band_structure}
\end{figure}

To understand the role of band structure on the results, we have studied the ground state properties for different values of the ratio of nearest- and next-nearest neighbor hopping matrix elements $t_2 / t_1$.  The non-interacting band structure is
\begin{equation} \label{eq:bandstructure}
\begin{split}
E(\vec{k})  = - 2 t_1 \Big[ \text{cos}(k_x) + \text{cos}(k_y) \Big]   \\
- 4 t_2 \Big[ \text{cos}(k_x) \text{cos}(k_y) \Big]  -\mu
\end{split}
\end{equation} 
where $\mu$ is the chemical potential.  For the 4-leg cylinder in the $L_x\to \infty$ limit, the (1D) Bloch wave-vector $k\equiv k_x$ takes on all values between $-\pi$ and $\pi$, while the discrete allowed values of $k_y$ serve as band-indices;  in the flux-free cylinder (periodic BCs), the allowed values of $k_y$ are $k_y = \{ 0, \pm \pi/2, \pi \}$ while for antiperiodic BCs (fl) $k_y = \{ \pm \pi/4, \pm 3\pi/4 \}$.  The corresponding band structures for periodic and antiperiodic BCs at $t_2 = 1/12$ are shown in Fig. \ref{fig:band_structure}, (a) and (b) respectively, where the horizontal red line indicates the Fermi energy for $\delta=1/8$.  The center band is doubly degenerate for periodic BC, and both  bands are doubly degenerate for antiperiodic BC.

In the interacting problem, as well, Bloch's theorem allows us to identify bands labeled by a band index, corresponding to the above values of $k_y$, and a 1D Bloch wave-vector, $k$, corresponding to $k_x$.  Although Fermi liquid theory does not apply in 1D, in some cases a Fermi surface can be identified from a calculation of the single-electron occupation probability, $\langle n_{\vec k,\sigma}\rangle$, if it has a non-analyticity  at $\vec k = k_F(k_y)\hat x + k_y\hat y$.  However, in practice it is difficult to distinguish a non-analyticity from a rapid analytic change.  Thus, as an operational definition of the Fermi momenta of the interacting system, we identify Fermi surface crossings with cases in which $\langle n_{\vec k,\sigma}\rangle$ drops sharply and the corresponding value of $k_F$ is associate with points of maximal slope, \emph{i.e.} points at which $d^2 \langle n_{\vec{k} \sigma} \rangle / d k^2 = 0$. A typical example is shown in Fig. \ref{fig:mom_kF} in the Supplementary Information.

A so-called ``generalized Luttinger's Theorem\cite{affleck_1997},'' which is really a generalized form of the Leib-Schulz-Mattis Theorem\cite{lieb_1961}, requires that in the absence of symmetry breaking, there must be a gapless neutral mode of the system with wave-vector $Q_{Lutt} \equiv 4\pi(1-\delta)$, which can be thought of as the almost Goldstone mode of an incommensurate CDW.  In contrast with Luttinger's theorem for Fermi liquids in more than 1D, this construction makes no direct reference to the single-particle spectrum at all.  However, if this could be interpreted in terms of fermiology, it would correspond to the ``total'' value
\begin{equation}
\begin{split}
k_F^{eff}=\sum_{k_y} k_{F} (k_y)
\rightarrow   \frac{Q_{Lutt}}{2} = 2 \pi (1 - \delta)  \quad 
\end{split}
\label{kFeff}
\end{equation}

\begin{figure}
  \centering
  \includegraphics[width=3.5in]{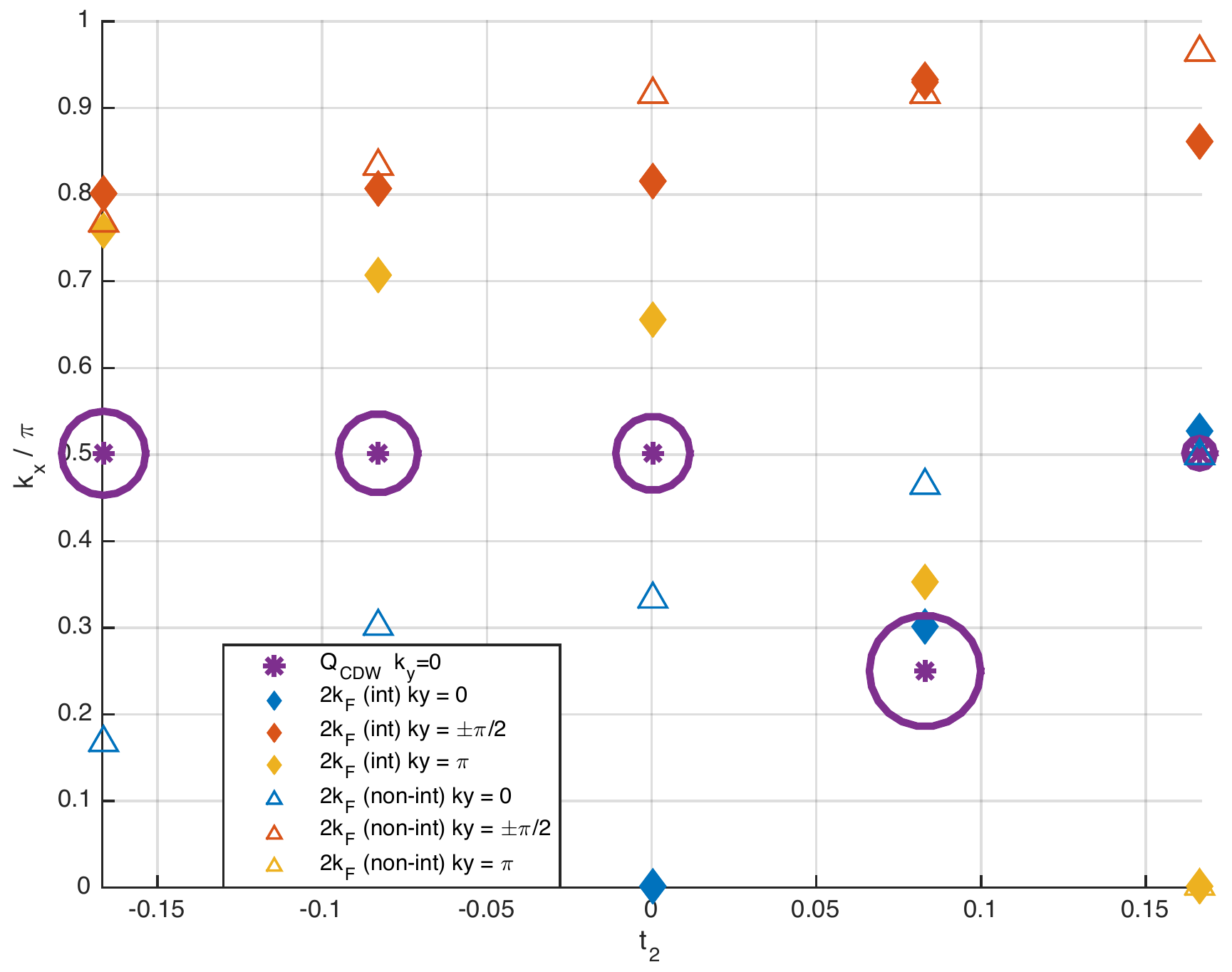}
  \caption{Charge ordering vectors and $2k_{F_i}$ for the $t$-$J$ model with $J_2 = 0$, $L_x = 48$, and periodic boundary conditions at different $t_2$.  Open triangles and filled diamonds represent non-interacting and interacting $2k_{F_i}$ respectively.}
  \label{fig:2kF}
\end{figure}

To study the possible origin of density wave order through Fermi surface nesting, we compare the charge ordering vector $Q_{cdw}$ to the non-interacting and interacting $2k_{F}(k_y)$ for each band.  The values of $Q_{cdw}$ are compared with the values of $2k_F$ in Fig. \ref{fig:2kF}.  The non-interacting $2k_{F}$'s are represented by open triangles and the interacting $2k_{F}$'s by filled diamonds.  We see that the ordering wave vectors do not coincide with $2k_{F_i}$ for each band which rules out CDW induced by Fermi surface nesting in our system.  Similar results are found across the $t_2 - J_2$ phase diagram suggesting another mechanism is responsible for the observed phases.  Additionally the non-interacting $k_F$ do not agree with the interacting $k_F$ showing a weak coupling perspective is not applicable to our results. 
 
We had hoped to identify cases in which this inference is violated in such a way that $k_F^{eff} = (4\pi/2) \delta$;  this, in turn, might have been suggestive of a tendency toward an exotic topologically order Fermi liquid phase (known as FL*), such as can occur\cite{sachdev_2003, sachdev_2015, white_2016} in certain versions of the quantum dimer model.\cite{rokhsar}  However, wherever well defined Fermi surface crossings can be identified in our data, we always find that they are suggestive of the Luttinger relation in Eq. \ref{kFeff}.

\section{Superconducting Correlations}
\label{superCorr}

\subsection{Charge $2e$}

We study charge 2$e$ superconducting order by measuring the pair-field correlation function $\Phi_{a, a^\prime}(\vec{r},\vec{r}^{\ \prime})$ as defined in Eq. (\ref{eq:SCPairField}).  We find that $\Phi_{x,x}$, $\Phi_{y,y}$, $\Phi_{x,y}$, and $\Phi_{x+y,x+y}$ all decay exponentially although we focus on $\Phi_{y,y}$ as it tends to have the largest amplitude.  We do not find quasi-long-range superconducting order in our study; however, we can still make qualitative statements about the relation between competing orders.  As noted earlier, increasing $t_2$ results in a weakening of charge order while maintaining N\'{e}el or stripe antiferromagnetic order depending on $J_2$.  The $J_2 = 0$ superconducting correlations increase slightly from $\xi_{sc} \approx 5.5$ to $\xi_{sc}\approx 6.7$ as $t_2$ increases from 0 to 1/6.  This enhancement of superconducting correlations with increasing $t_2$ (accompanied with weakening of CDW) suggests a competition between superconductivity and charge order.  We find pronounced amplitude oscillations in the superconducting correlation functions with the same period as the CDW which will be discussed below.

Compared with the periodic boundary condition, there is a clear enhancement of charge $2e$ superconducting correlations for the case with antiperiodic boundary conditions (flux).  We find the largest correlation lengths for $J_2 / J_1 = 0.3$ \& $0.46$ corresponding to the quantum paramagnet with no spin and charge order (Liquid II phase). For these values of $J_2$, the superconducting correlations appear more uniform with a correlation length of $\xi_{sc}\sim 10$ for $J_2/J_1=0.3$ and $t_2=1/6$ shown in Fig. \ref{fig:supercor} (a).  This enhancement of uniform superconducting correlations coinciding with the destabilization of stripes (by flux and magnetic frustration) serves as additional evidence for a competition between these orders.

\subsection{Pair Symmetry} \label{pair_symmetry}

\begin{figure}
  \centering
  \includegraphics[width=2.8in]{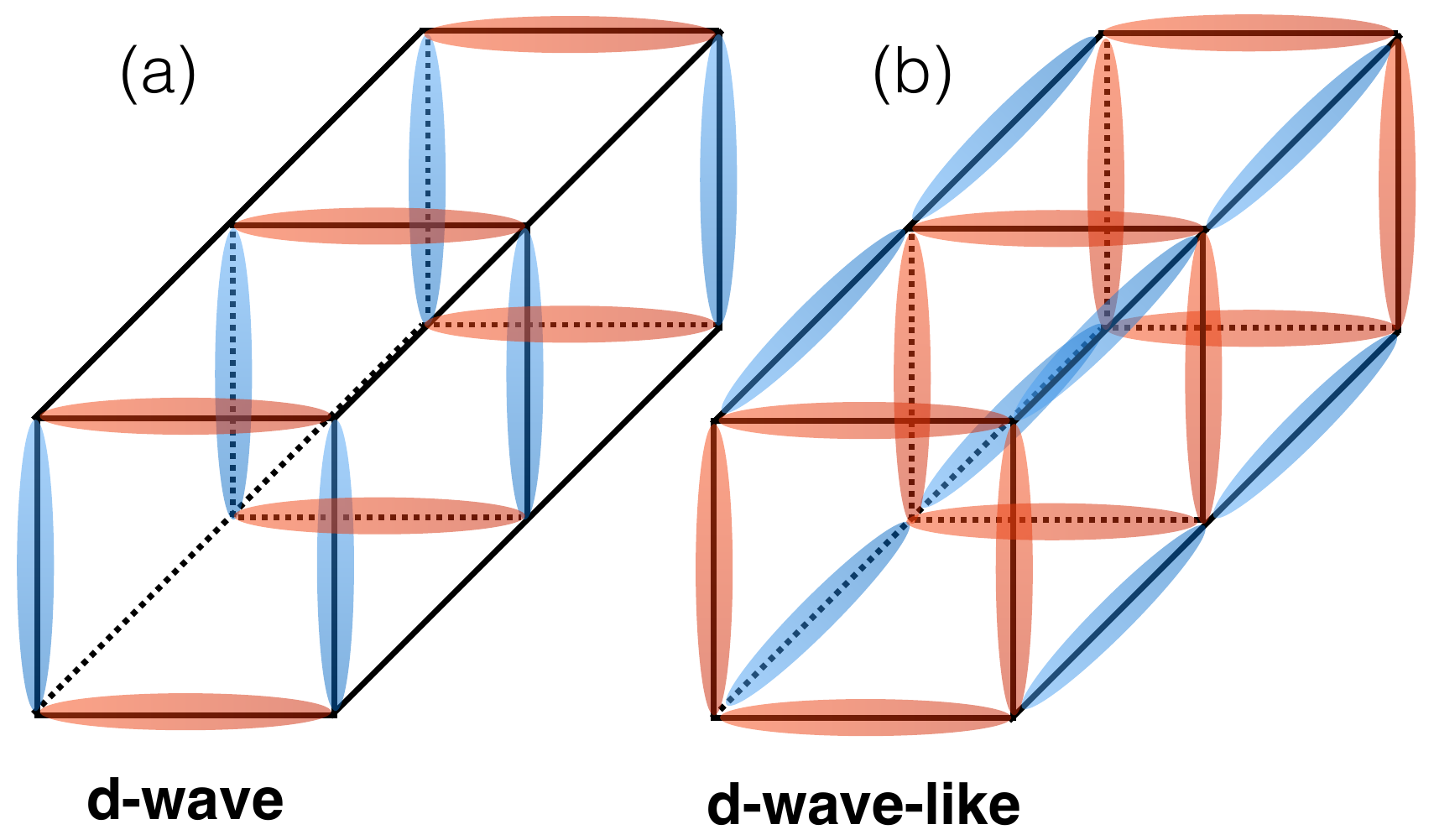}
  \caption{Superconducting correlations may have (a) $d$-wave symmetry characterized by a change in sign (color) of the order parameter \emph{around} the cylinder or (b) $d$-wave-like which has $s$-wave symmetry around the cylinder, but changes sign about plaquettes along the cylinder.}
  \label{fig:d-wave-like}
\end{figure}

Although the locally $d$-wave character of the superconducting correlations has long been noted in DMRG studies of $t$-$J$ ladders,\cite{white_1997, white_1999}  in the cylinder geometry a precise $d$-wave symmetry can be defined with respect to the $C_4$ rotational symmetry about the axis of the cylinder.  Indeed, looking at the long-distance behavior of the pair-field correlation function, we distinguish two cases, as shown schematically in Fig. \ref{fig:d-wave-like}: correlations can be said to be ``$d$-wave'' if they are odd under this $C_4$ rotation, and ``$d$-wave-like''  if they are even under rotation, but have opposite signs on the $x$ and $y$ directed bonds.  The long-distance behavior of the correlation function corresponds to true $d$-wave symmetry everywhere in the phase diagram except in the CDW-4d, Liquid I, and Liquid II.  Of these, the first two show clear $d$-wave-like symmetry, while Liquid II cannot be easily classified by symmetry at all.  Note that these are the phases which have weak or vanishing charge order. 
 
\begin{figure}
  \centering
  \includegraphics[width=3.45in]{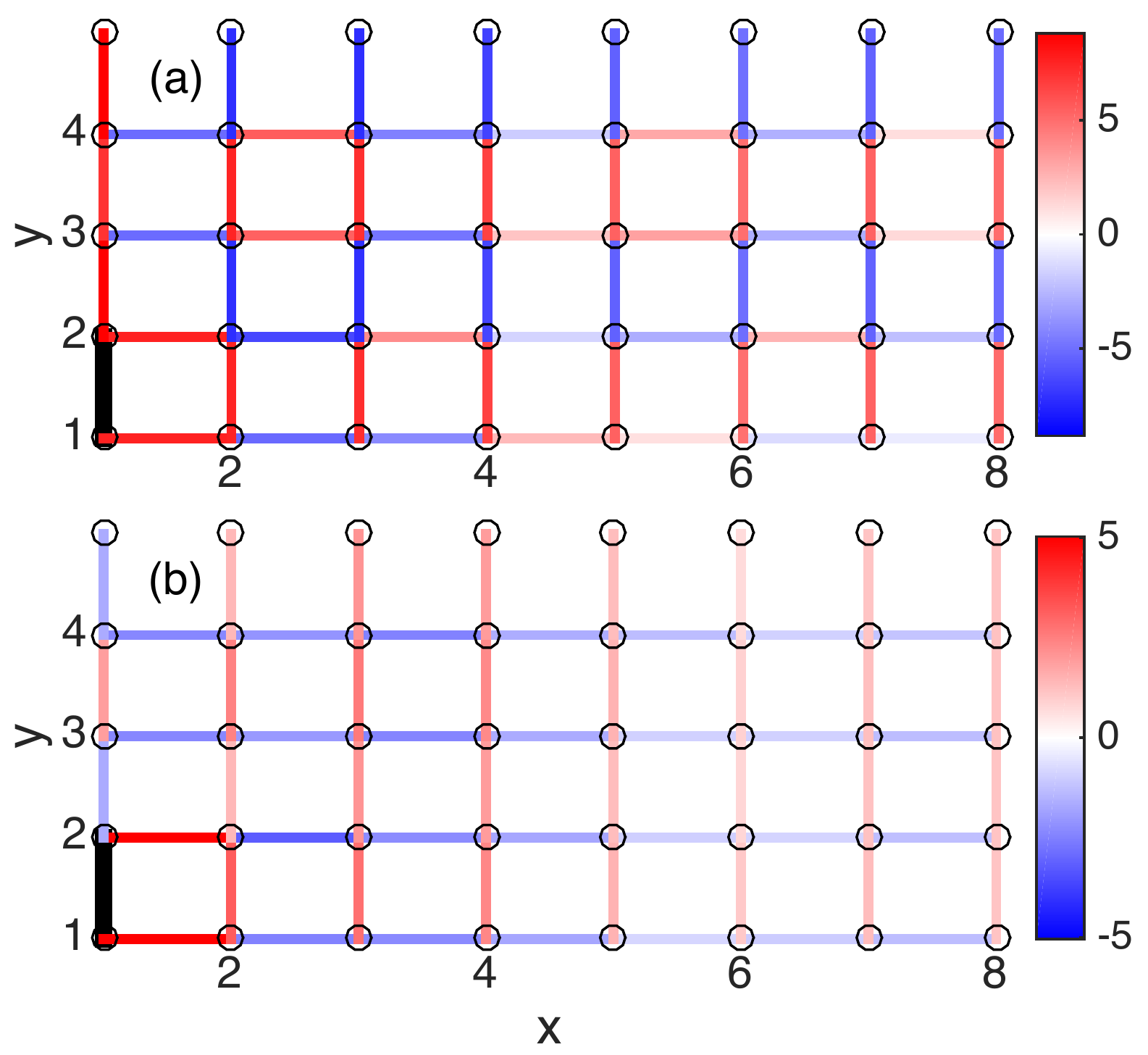}
  \caption{Superconducting correlation functions on the cylinder for (a) CDW-4a phase with $t_2 = J_2 = 0$ and (b) CDW-4d phase with $t_2 = 1/6$ and $J_2 = 0$.  The black line denotes the reference $y$ bond and vertical (horizontal) colored lines denote $\log [1 + |\Phi_{y,y}| / |\Phi_0| ]$ ($\log [ 1 + |\Phi_{y,x}| / |\Phi_0| ]$) where $\Phi_0$ is the minimum of the correlation function. Red (blue) is used where the correlation function is positive (negative).  The change in color around the cylinder shows the sign change of the long-distance superconducting correlations which suggests $d$-wave symmetry in the CDW-4a phase and $d$-wave-like symmetry in the CDW-4d phase.}
  \label{fig:d-wave}
\end{figure}

The $\Phi_{y,y}$ ($\Phi_{y,x}$) correlation function is shown in Fig. \ref{fig:d-wave} along the $y$ ($x$) bonds for (a) the CDW-4a phase ($t_2 = J_2 = 0$) and (b) the CDW-4d phase ($t_2 = 1/6$ and $J_2 = 0)$ of the $t$-$J$ model with respect to the vertical black bond at the bottom left corner. The long-distance sign pattern of the superconducting correlations demonstrates $d$-wave and $d$-wave-like symmetries matching Fig. \ref{fig:d-wave-like}.  We have confirmed this pattern by applying an external pair field $D_{\vec{a}}(\vec{r}_i)$ along one rung at the edge of the cylinder and measuring the response in the system bulk.

\subsection{Pair-field Oscillation}

The superconducting pair-field correlations show varying degrees of oscillation across the phase diagrams with and without flux.  The pair-field component $\phi_{\vec{Q}}$ at non-zero wave vector $\vec{Q}$ can couple to the CDW order parameter $\rho_{\vec{Q}}$ through terms in the energy of the form\cite{berg_2009, intertwined}:
\begin{equation} \label{eq:coupled}
\rho_{\vec{Q}} \phi^*_{\vec 0} \phi_{- \vec{Q}} + \text{c.c.}   \quad \quad \quad \& \quad \quad \quad \rho_{2 \vec{Q}} \phi^*_{-\vec{Q}} \phi_{-\vec{Q}} + \text{c.c.}
\end{equation}
such that when two of the factors in the first term are present, the third one is also expected to be present on general grounds of Landau theory.  Since the second term is quadratic in $\phi_{\vec{Q}}$, a pair-density wave (PDW) component will be accompanied by charge ordering with twice the wave vector, but the converse is not true.  This indicates that the PDW is a bona fide phase which can exist independently of other orders.  Observation of a superconducting component with half the CDW wave vector  would have suggested an intrinsic tendency towards PDW order.  However, we only find a density wave component in the correlation function $\Phi^{(dw)}$ with the same period as the CDW.

The pair-field correlations are enhanced when flux is threaded through the cylinder and exhibit oscillations in the amplitude of the superconducting correlation functions, for instance, for $J_2 = 0$ and $t_2 = 1/12$ as well as $J_2 / J_1 = 0.56$ \& $0.80$.  The period is again the same as the charge order suggesting a strong coupling between CDW and uniform SC fluctuations with correlations that can be fit to Eq. (\ref{eq:phi_PDW}).  We find values of $\Phi^{(dw)}_{y,y} / \Phi^{(sc)}_{y,y} \sim 0.11 - 0.97$ with the largest values occurring for larger $J_2 / J_1$, e.g. $J_2 / J_1= 0.56$ \& $0.80$ (see Fig. \ref{fig:supercor} (b) and Fig. \ref{fig:contour} in Supplementary Information) corresponding to (weak) period 8 charge order and short-ranged $[0,\pi]$ spin order, while the pair-field oscillations are smaller for other parameters (see Fig. \ref{fig:supercor} (a)).  While we do not find a finite PDW component at $Q_{cdw}/2$, the observation of an induced density wave $\Phi^{(dw)}$ comparable to the uniform superconducting component $\Phi^{(sc)}$ allows us to conclude that the coupling of the first term in Eq. (\ref{eq:coupled}) is significant in our model.  We find $\Phi^{(dw)} / \Phi^{(sc)}$ tends to increase with larger $t_2$ when flux is threading the cylinder, which is an expected consequence of Eq. (\ref{eq:coupled}).

\begin{figure}
  \centering
  \includegraphics[width=3.43in]{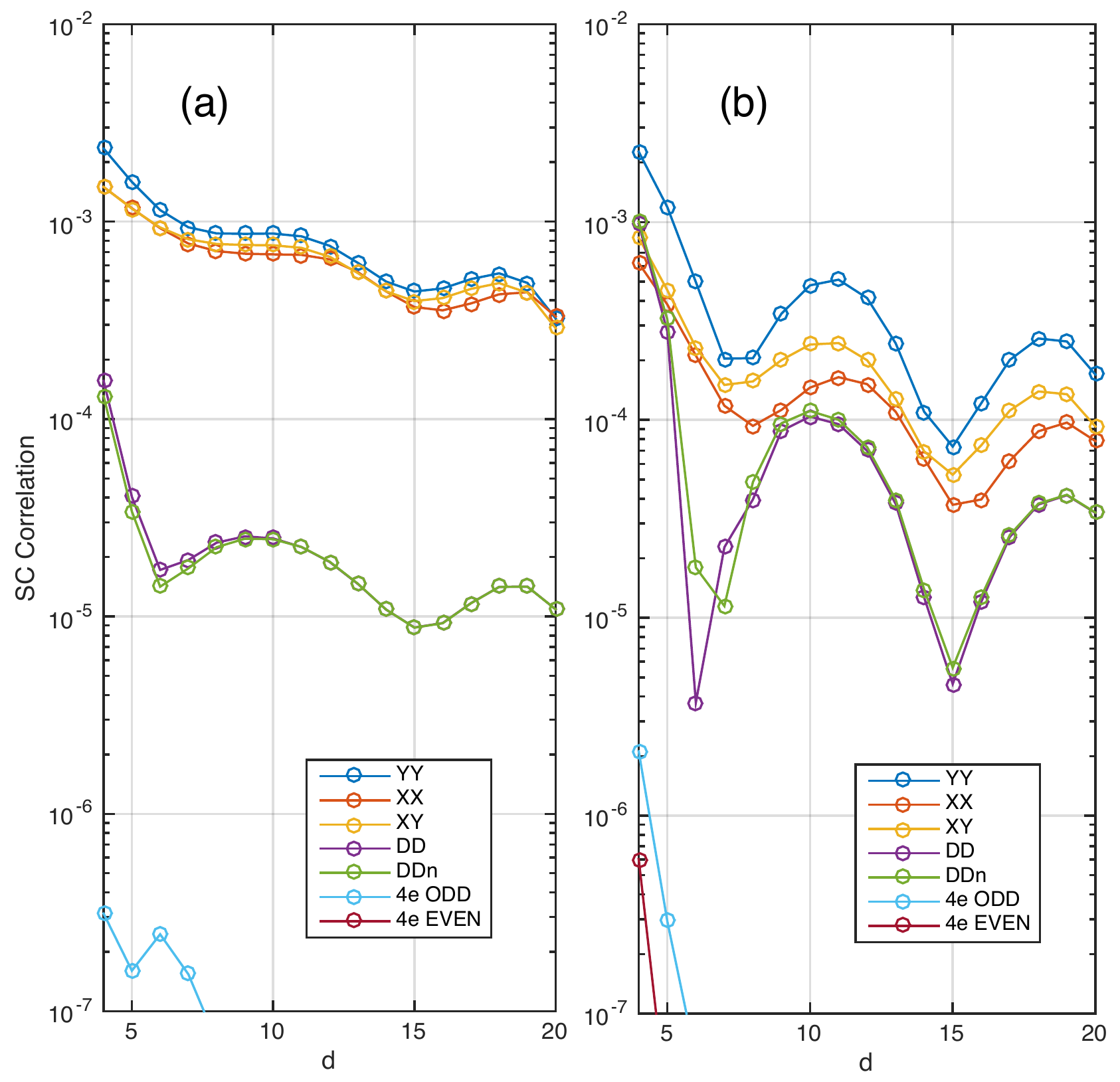}
  \caption{Superconducting correlations for the $t$-$J$ model with flux as a function of separation $d$ along $\hat{x}$ for (a) $J_2 = 0.3 J_1$ and $t_2 = 1/6$ showing strong uniform component $\Phi^{(sc)}_{y,y}$ and (b) $J_2 = 0.56 J_1$ and $t_2 = 1/6$ showing a significant induced density wave component $\Phi^{(dw)}_{y,y}$.}
  \label{fig:supercor}
\end{figure}

\subsection{Charge $4e$}

We have  also checked the correlations of the two singlet ``quartet'' charge $4e$ operators $\phi_{4e,\pm}$ as defined in Eq. (\ref{eq:SC4eEVEN}) and (\ref{eq:SC4eODD}).  The correlation function is found to decay rapidly and fall to below numerical error within a couple lattice constants across the entire phase diagram with and without flux (as can be seen, for example, in Fig. \ref{fig:supercor}).   Charge $4e$ superconductivity seems unlikely in the present model.

\section{Energy Gaps} \label{energyGap}

\begin{figure}
  \centering
  \includegraphics[width=3.5in]{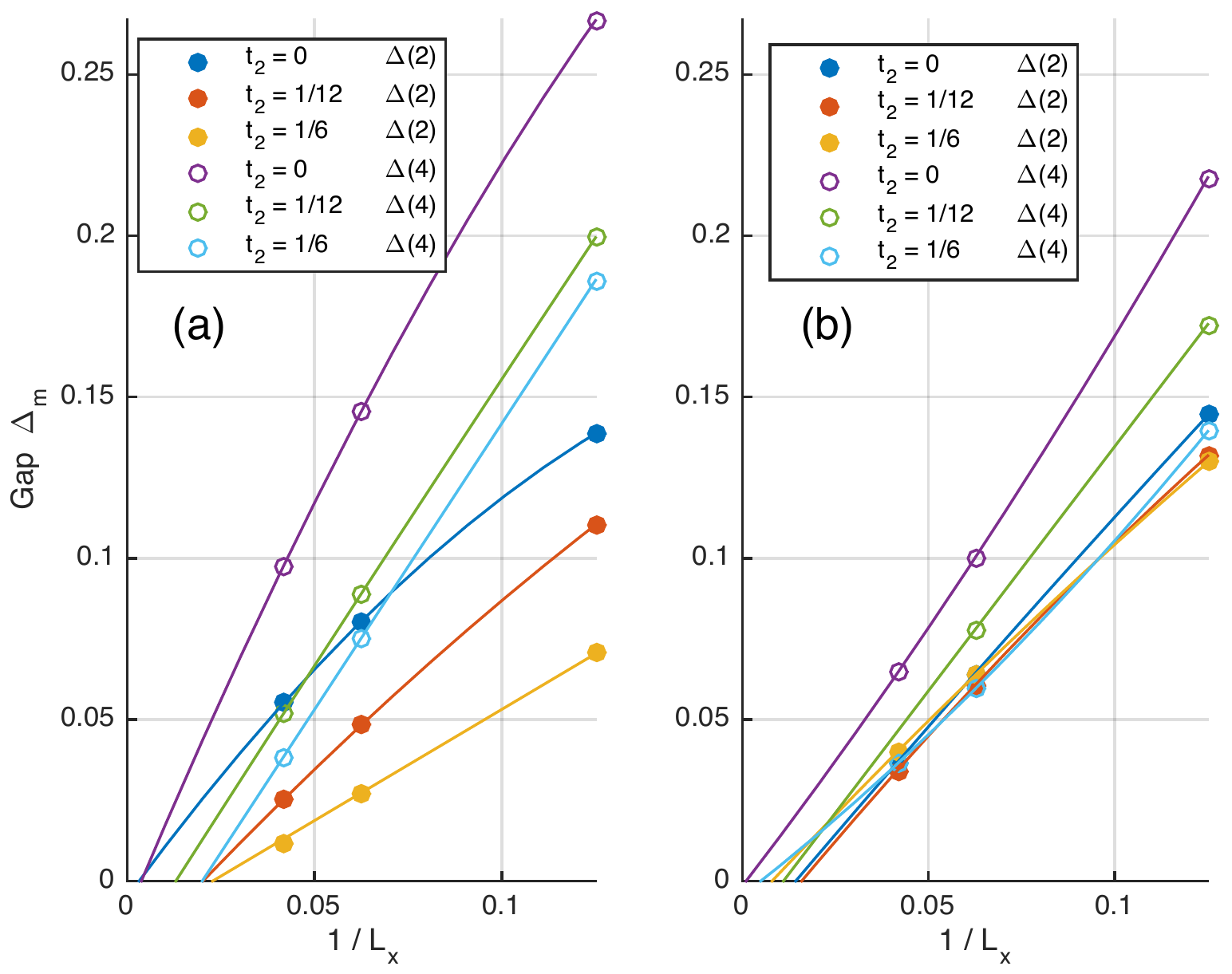}
  \caption{Energy gap $\Delta_m$ defined in Eq. (\ref{Eq:EnergyGap}) of the $t$-$J$ model under (a) periodic and (b) antiperiodic boundary conditions at $J_2 = 0$ and different $t_2$.}
  \label{fig:gapM}
\end{figure}

Energy gaps can be determined by adding particles to the system and comparing the ground state energies.  We define the energy gap $\Delta_m$ as the energy (per particle):
\begin{equation} \label{Eq:EnergyGap}
\Delta_m = \frac{1}{m} \Big[ E_0(N+m) + E_0(N-m) - 2 E_0(N)  \Big] ,
\end{equation}
where $E_0(N)$ is the ground state energy of the system with $N$ particles.  For a charge $2e$ superconductor, there is a gap to add one particle, but no gap to add two or four particles, \emph{i.e.} $\Delta_1 \neq 0$ and $\Delta_2 =  \Delta_4 = 0$.  We compute these gaps for $L_x\times 4$ cylinders with $L_x = 8, 16, 24$ and extrapolate to $L_x \rightarrow \infty$.  The gaps $\Delta_2$ and $\Delta_4$ computed from the DMRG simulations are plotted in Fig. \ref{fig:gapM} for $J_2 = 0$ and $t_2 = \{ 0, 1/12, 1/6 \}$ (a) without and (b) with flux.  The single particle gap $\Delta_1$ for $J_2 = 0$ (not shown) is largest for $t_2 = 0$ while close to zero for $t_2 = 1/6$ where the CDW order is weaker for larger $t_2$.  The opposite is true for the case with flux:  the CDW order and single particle gap $\Delta_1$ are largest for $t_2 \sim 1/6$.  Therefore, our results suggest the presence of Cooper pairing corresponding to the finite energy gap $\Delta_1$.  

However, we find $\Delta_2$ and $\Delta_4$ appear to vanish upon extrapolation.  This seemingly is at odds with the observed finite superconducting correlation lengths $\xi_{sc}\lesssim 10$.  

\section{Discussion}
\label{discussion}

\subsection{Intertwined Orders}

Despite the simplicity of the model we have studied, we have unveiled a remarkably complex intertwining between at least 9 forms of order (CDW-4, CDW-8, SDW-8, SDW-16, SDW $0-\pi$, SDW $\pi-0$, SDW $\pi-\pi$, $d$-wave and $d$-wave-like SC).  There is a general tendency for terms which depress CDW order to enhance the SC correlations, showing that there is some form of ``competition'' between these two orders.  Where the superconducting correlations show oscillatory spatial structures,  the period is always  the same as that of the CDW, and hence can also be understood in terms of a competition between CDW and uniform SC orders.  CDW induced oscillations of the SC order parameter similar to what we have observed have recently been imaged on the surface of Bi2212 directly by Josephson tunneling microscopy\cite{hamidian_2016}, and were inferred previously from features of the quasiparticle spectrum in scanning tunneling microscopy.\cite{aharon_2003}

The relation between SDW and CDW order, by contrast, is less obvious.  For small $t_2$ and no flux, we see robust period 4 CDW order.  For small $J_2/J_1$, there are corresponding period 8 SDW correlations, but upon increasing $J_2/J_1$ to $\sim 1/2$, the magnetic correlations essentially disappear.  The interplay between spin and charge order can produce diverse macroscopic realizations.\cite{zaanen_2001}  This observation offers a possible connection between the intertwined order in LSCO, lack of substantial static spin correlations in BSCCO, and apparently unrelated charge and spin ordering in YBCO.  The relative insensitivity of the charge ordering to the degree of spin ordering apparent in our study supports the perspective that the charge density waves which have been observed across different families of underdoped cuprates have a common origin.

\subsection{Origin of the Density Wave Orders}

We have studied the role of Fermi surface nesting in density wave formation on the 4-leg cylinder.  Since nesting is perfect in 1D and problematic in higher dimensions, the cylinder geometry we have studied represents the strongest candidate for observing nesting-induced CDW.  The fact that we do not find a clear relationship between $Q_{cdw}$ and the values of $2k_{F}$ suggests Fermi surface nesting does not generally play a role in CDW formation.

From the strong coupling perspective, we can view the CDW-4a (half-filled stripe) ground state as a Wigner crystal of singlet pairs which delocalize around the cylinder.  The finite spin correlation length, $\xi_{sdw}$, can perhaps be thought of as the size of a Cooper pair, and the energy to break a pair can be associated with the single particle gap $\Delta_1$ to add a hole to the system.

\subsection{Doped Spin Liquid}

Given that the spin-1/2 antiferromagnetic square $J_1-J_2$ Heisenberg model exhibits\cite{jiang_2012, donnasheng, wen_2016} an intermediate quantum paramagnetic phase, and possibly contains a spin liquid sub-regime, we had initially hoped to see an interesting remnant of this when we doped the system, forming a ``doped spin liquid.''  Arguments were presented by Anderson\cite{anderson_1987}, and one of us\cite{kivelson_1987, rokhsar} that a spin-liquid can be viewed as a paired state with zero superfluid stiffness, so that upon doping it would inevitably form a strongly superconducting state.  More recently, it has been proposed that a doped spin liquid might form an  FL* phase, which in turn has been proposed as a candidate explanation of the physics of the pseudo-gap regime of the hole-doped cuprates.\cite{sachdev_2003, sachdev_2015, debanjan}  Unfortunately, CDW order seems to be the dominant ordering tendency, even in the doped spin liquid range of parameters, and no evidence of either of these conjectured behaviors has yet been seen in our studies.

\subsection{Superconducting Symmetry}

It is an interesting feature of the cylinder geometry that a true symmetry distinction between d-wave and s-wave superconducting states is possible.  The superconducting correlations in our model were shown to have $d$-wave symmetry across the phase diagram where charge order is present, and $d$-wave-like symmetry (although this is technically s-wave symmetry with respect to rotation about the axis of the cylinder) where charge order is absent or weak. 

For moderate $J$, particles on a single plFD was supported aquette tend to pair;  we can therefore imagine a stack of plaquettes, each favoring one or two pairs where the 2-electron ground state has $s$-wave symmetry and the 4-electron ground state has $d$-wave symmetry.\cite{trugman_1996}  Turning on a coupling between plaquettes, pair hopping then corresponds to a bosonic operator with $d$-wave symmetry which manifests in a sign change of the superconducting correlation function upon rotating the cylinder by $\pi/2$.  

Away from $\delta=1/8$, and especially for $\delta \sim 0.15$, where the CDW ordering tendencies are weaker, in agreement with other studies of $t$-$J$ ladders,\cite{white_1997} we observe substantially enhanced SC correlations, as we will report in a separate communication.

\section*{Acknowledgements}

We would like to thank E. Berg and J. Tranquada for helpful discussions.  HCJ was supported by the Department of Energy, Office of Science, Basic Energy Sciences, Materials Sciences and Engineering Division, under Contract DE-AC02-76SF00515. SAK was supported in part by NSF grant \#DMR 1265593 at Stanford.  The computational work was partially performed at Sherlock cluster in Stanford.


\appendix 

\begin{center}
\noindent {\large {\bf Supplementary Information}}
\end{center}

\renewcommand{\thefigure}{S\arabic{figure}}
\setcounter{figure}{0}
\renewcommand{\theequation}{S\arabic{equation}}
\setcounter{equation}{0}

\section{Long-Range Charge Order}

\begin{figure}
  \centering
  \includegraphics[width=3.4in]{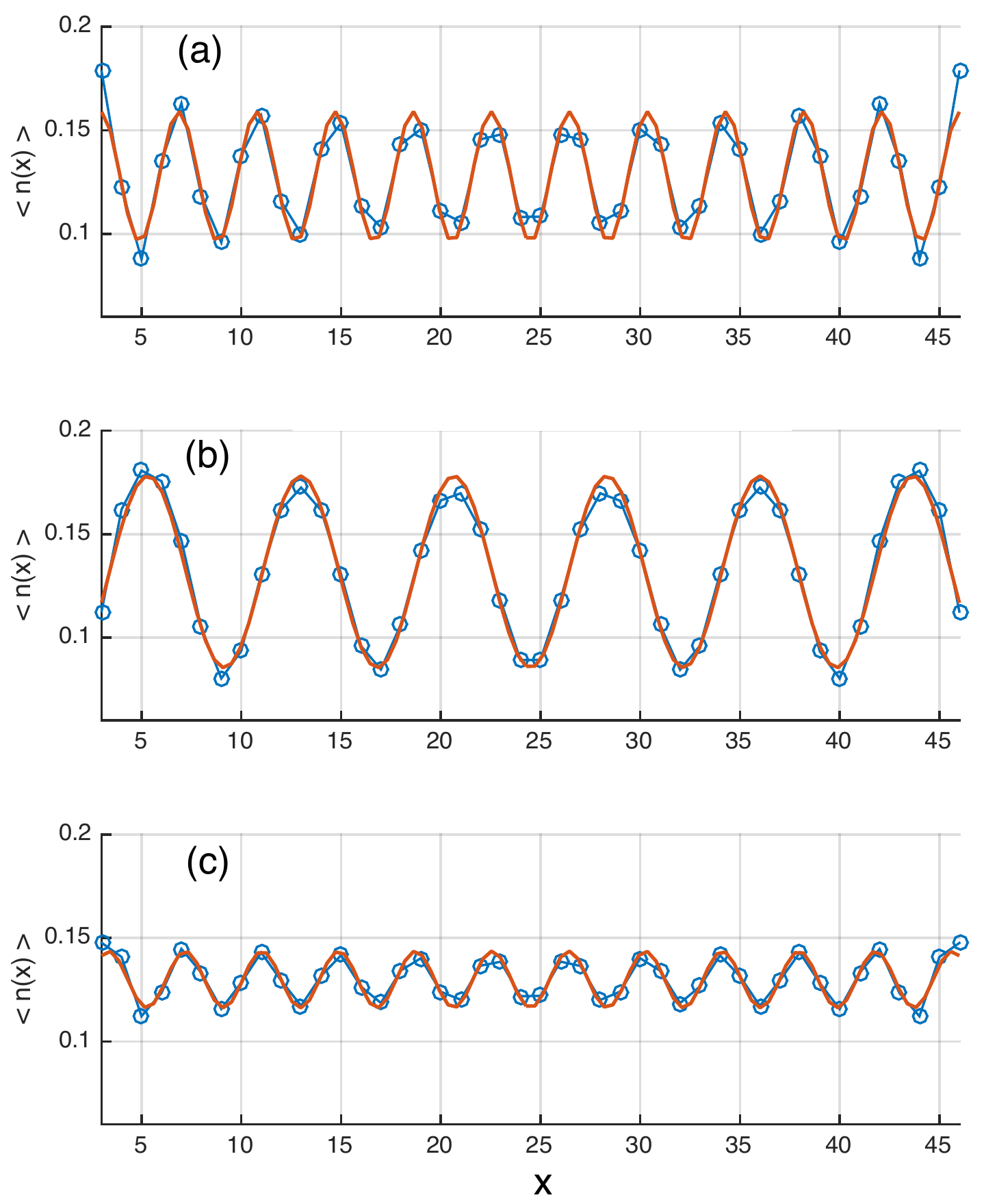}
  \caption{The charge density profile $\langle n(x)\rangle$ for the $t$-$J$ model without flux at $J_2=0$, and (a) $t_2 = 0$, (b) $1/12$, and (c) $1/6$ with $L_x = 48$. The open circles represent the raw numerical data, while the red lines are fitting curves using function $n(x)=\rho_{cdw} \cos(Q_{cdw}\cdot x + \theta) + A$, where $\rho_{cdw}$ and $Q_{cdw}$ are CDW order parameter and ordering wave vector, respectively. Two points were removed from both ends in the fitting in order to minimize boundary effects.}
  \label{fig:amp_fit}
 \end{figure}

We have determined the existence of long-range CDW order in the 4-leg $t$-$J$ cylinders by fitting the amplitude $\rho_{cdw}$ of the oscillation of the density profile $\langle n(x) \rangle$ in the system and extrapolating the value to $1 / L_x \rightarrow 0$. The fit for $J_2 = 0$, $t_2 = \{ 0, 1/12, 1/6 \}$, and $L_x = 48$ is shown in Fig. \ref{fig:amp_fit} where we have removed two points from both ends to minimize boundary effects.  

\begin{figure}
  \centering
  \includegraphics[width=3.4in]{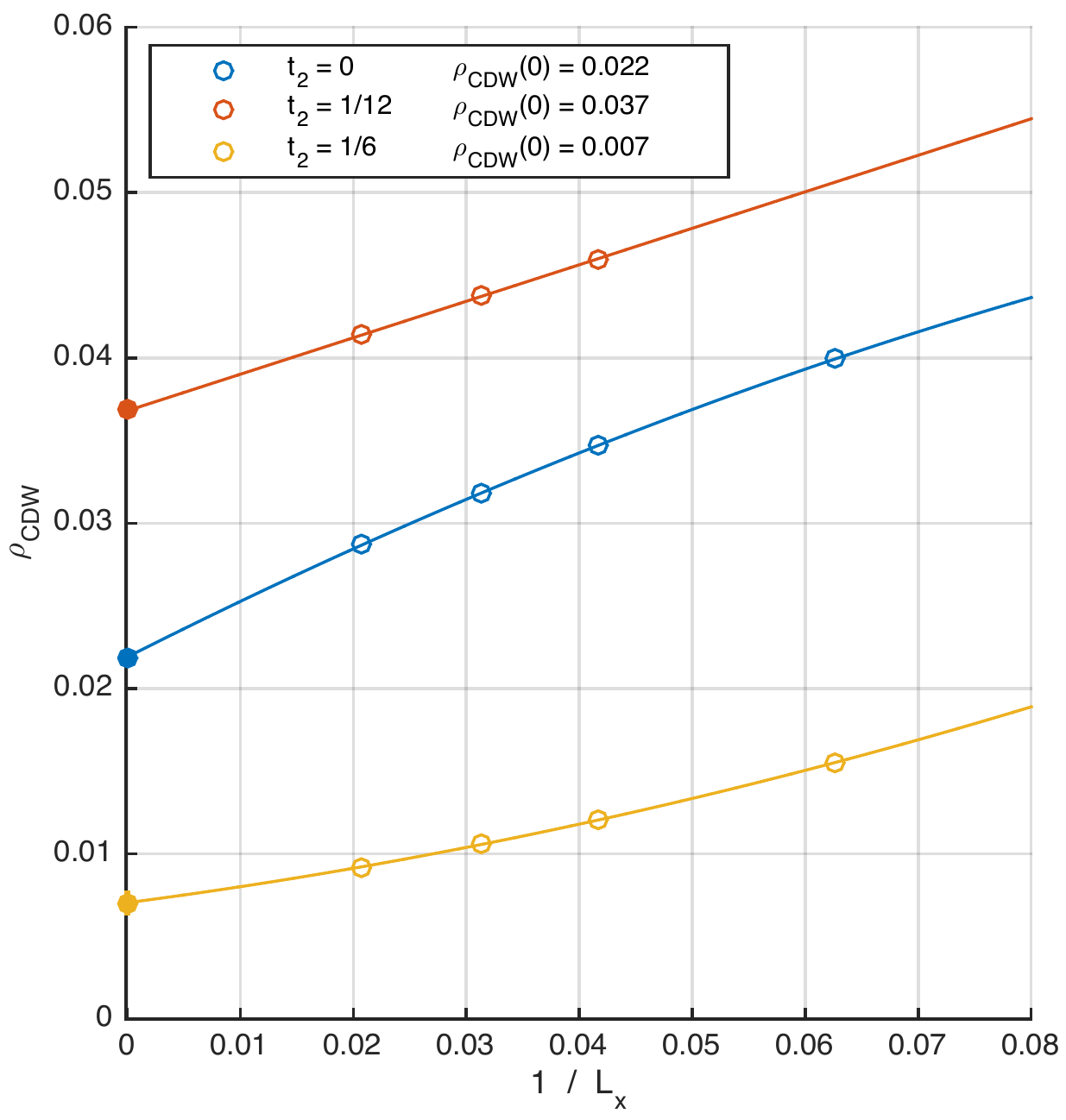}
  \caption{Amplitude $\rho_{cdw}$ of oscillation in the density profile $\langle n(x) \rangle$ for the $t$-$J$ model at $J_2 = 0$ and $t_2 = 0$ , $1/12$ and $1/6$ (open circles) and its extrapolation to $1 / L_x \rightarrow 0$ (filled circles) fitted by quadratic polynomial function labelled by the solid lines.}
  \label{fig:scale_j2-0}
\end{figure}

An example of the extrapolation for $J_2 = 0$ and $t_2 = \{ 0, 1/12, 1/6 \}$ is shown in Fig. \ref{fig:scale_j2-0} for system sizes $L_x = 16, 24, 32, 48$, where the solid circle denotes the amplitude in the $L_x=\infty$ limit. The CDW-4a (blue), CDW-8a (red), and CDW-4d (yellow) extrapolate  to non-zero values with quadratic polynomial fit (error bars correspond to 95\% confidence interval, or $\sigma = 1.96$ standard deviations).

\section{Determining $k_F$}

\begin{figure}
  \centering
  \includegraphics[width=3.45in]{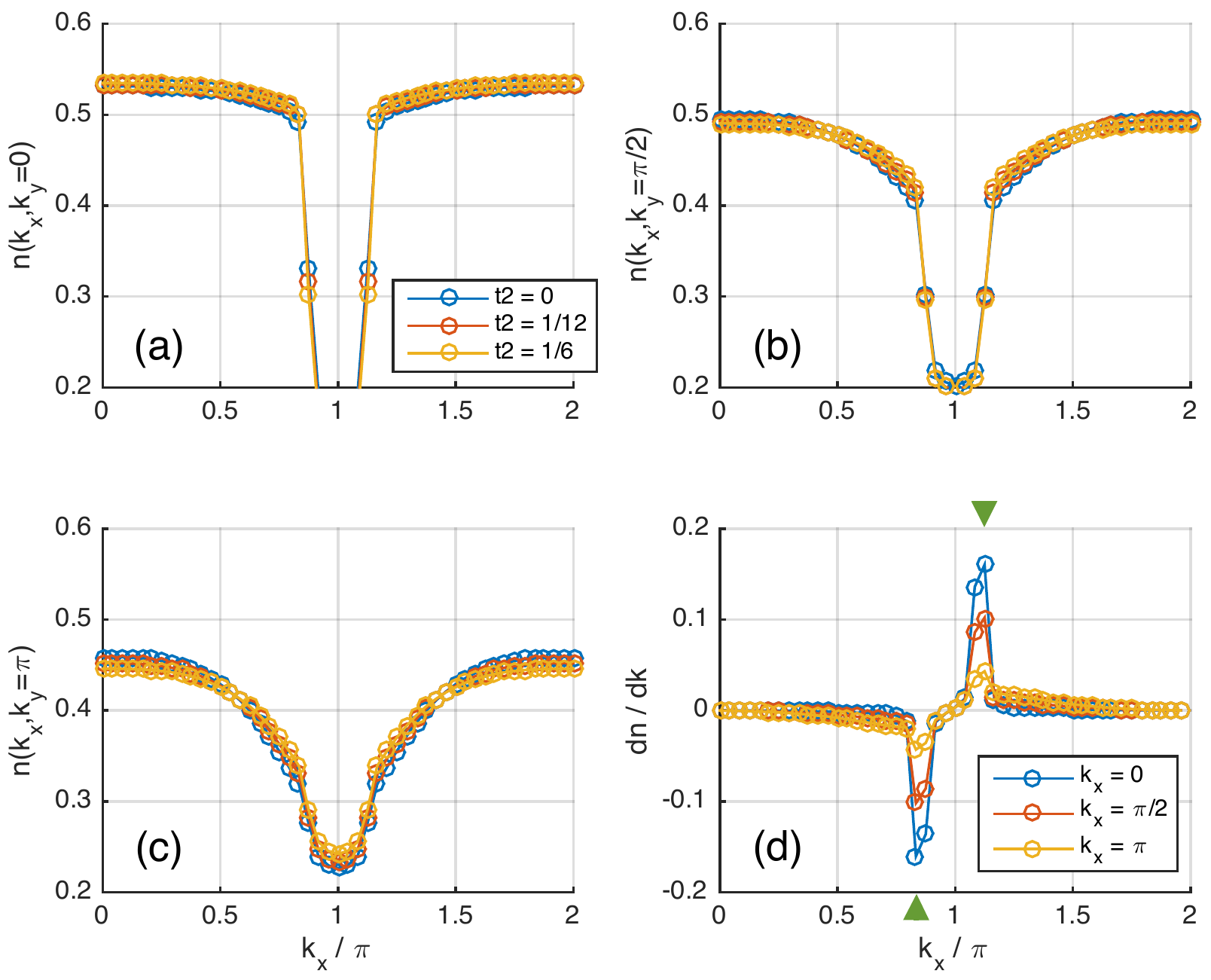}
  \caption{Single-electron occupation probability $\langle n_{\vec{k},\uparrow} \rangle$ for $k_y = 0, \pm\pi/2, \pi$ bands. A sharp drop can be identified for $J_2 / J_1 = 0.80$ with antiperiodic boundary conditions at (a) $t_2 = 0$, (b) $t_2 = 1/12$, and $t_2 = 1/6$. The first derivative for (d) $t_2 = 0$ for $k_y = \{ 0, \pi/2, \pi\}$ shows a clear peak (green arrow) at $k_F(k_y) \sim 7/8$.}
  \label{fig:mom_kF}
\end{figure}

We determine $k_F(k_y)$ for each $k_y$ in the interacting problem from the single-electron momentum distribution function $\langle n_{\vec{k},\sigma} \rangle$. As can be seen in Fig. \ref{fig:mom_kF}, a clear drop can be identified for the $t$-$J$ model with antiperiodic boundary conditions at $J_2 / J_1 = 0.80$ for (a) $t_2 = 0$, (b) $t_2 =1/12$, and (c) $t_2 = 1/6$. This clear drop or jump in $\langle n_{\vec{k},\sigma} \rangle$ correspond to a sharp peak in $d \langle n \rangle / d k_x$ in (d), where the peak position labels $k_F(k_y)=k_x(k_y) \sim 7\pi/8$ for $k_y = \{ 0, \pi/2, \pi\}$ at $t_2 = 0$. The interacting values of $2 k_F$ can then be compared with the non-interacting band structure as well as the charge and spin ordering vectors to understand the role of Fermi surface nesting.

\section{Loop Current}

\begin{figure}
  \centering
  \includegraphics[width=3.4in]{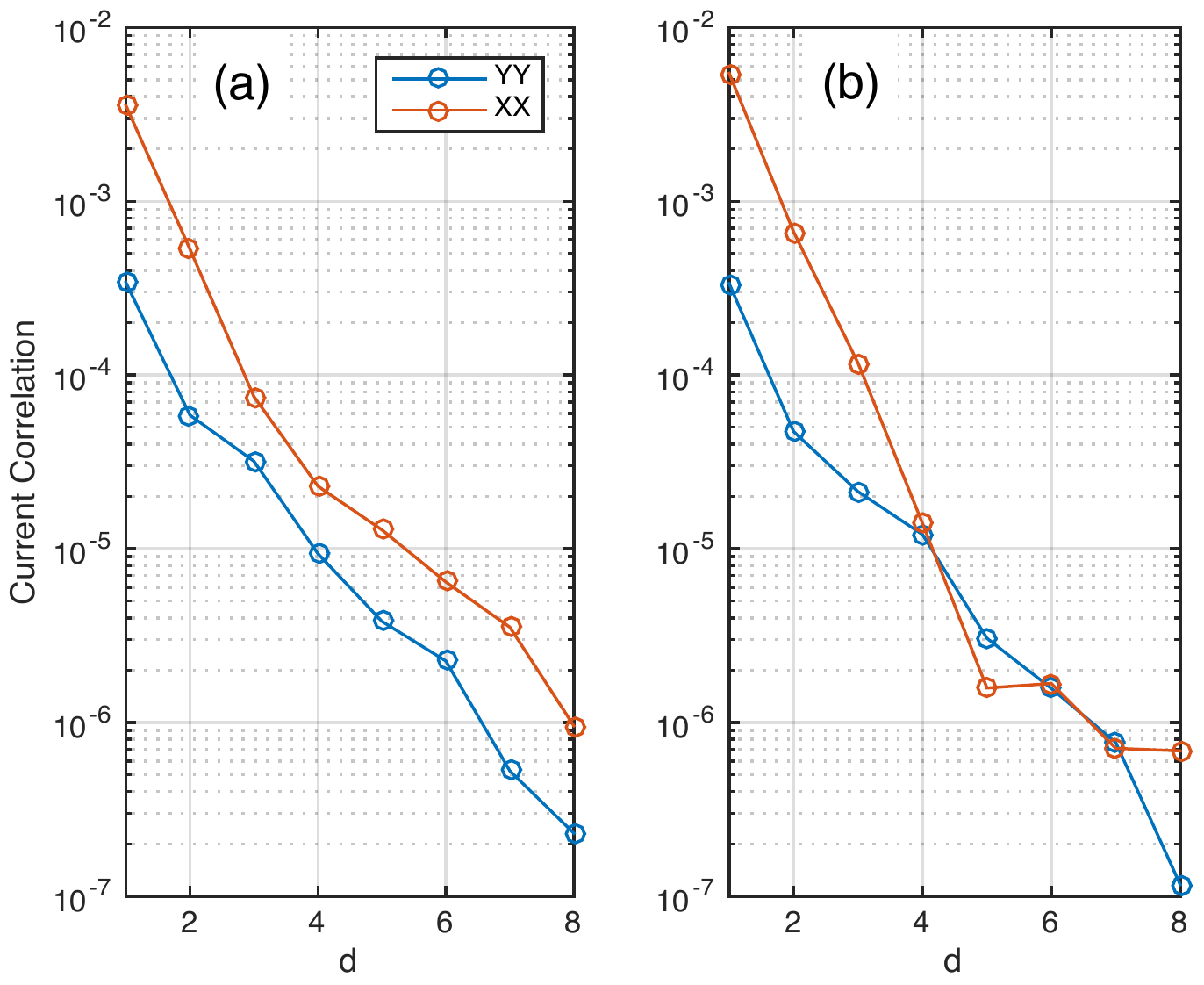}
  \caption{Current correlation function for $J_2 = 0$ and (a) $t_2 = 1/12$ and (b) $t_2 = 1/6$ for bonds along $\widehat{x}$ (red) and $\widehat{y}$ (blue).}
  \label{fig:loop}
\end{figure}

We searched for loop current order by studying the behavior of the current-current correlation function and found exponential decay across the phase diagrams for the $t$-$J$ model with periodic and antiperiodic boundary conditions. Fig. \ref{fig:loop} shows the current-current correlations for $J_2 = 0$ and (a) $t_2 = 1/12$ and (b) $t_2 = 1/6$ showing rapid exponential decay with the separation $d$. The correlation lengths $\xi_{loop} < 2 = L_y / 2$ implying current correlations are not significant in our model for the range of parameters chosen.

\section{Pair-Field Strength}

\begin{figure}
  \centering
  \includegraphics[width=3.2in]{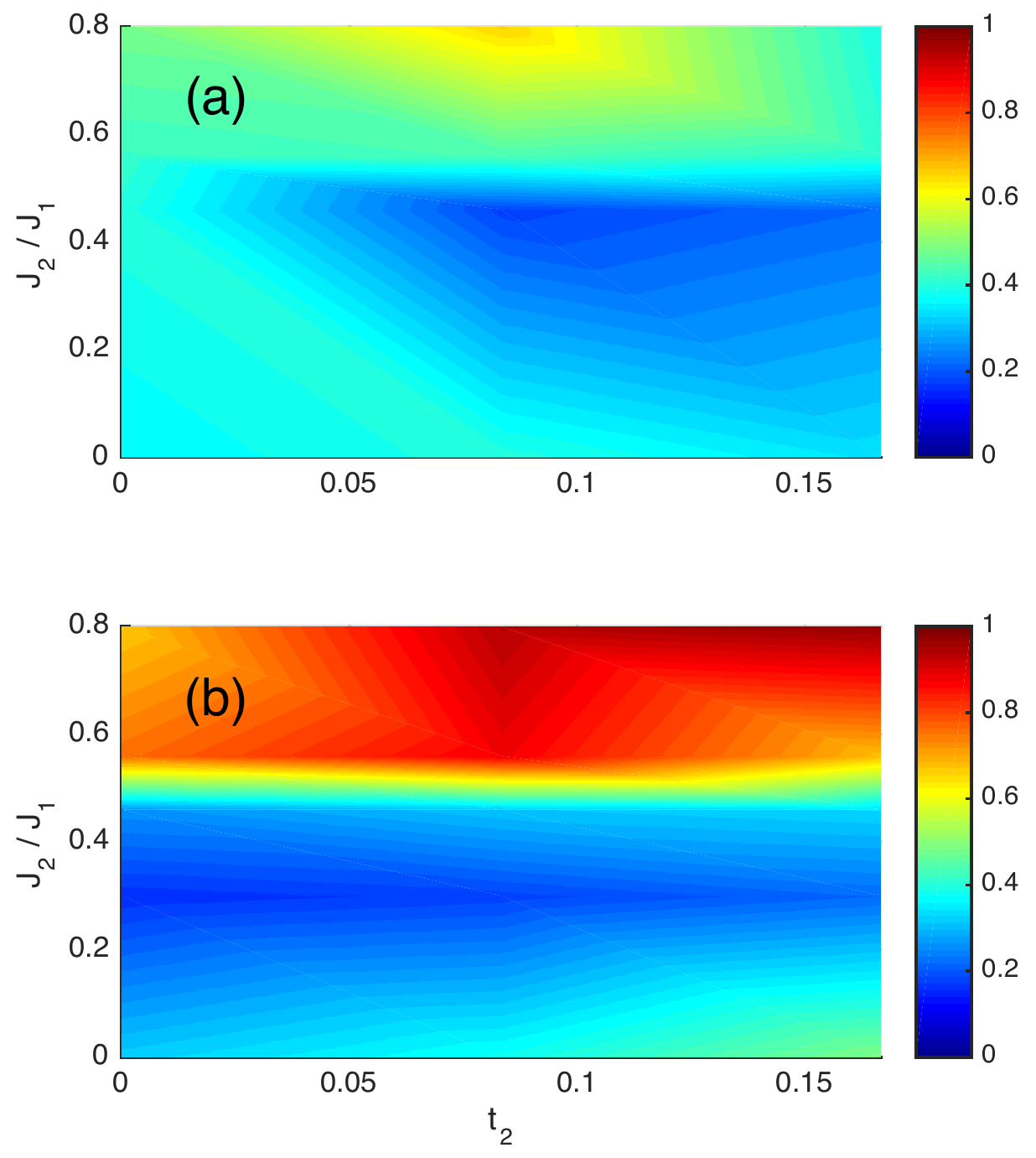}
  \caption{Plot of $\Phi^{(dw)}_{y,y} / \Phi^{(sc)}_{y,y}$ as functions of $t_2$ and $J_2 / J_1$ with (a) periodic and (b) antiperiodic boundary conditions around the cylinder.}
  \label{fig:contour}
\end{figure}

The strength of the pair-field correlation function at finite wave vector is shown in Fig. \ref{fig:contour} where $\Phi^{(dw)}_{y,y} / \Phi^{(sc)}_{y,y}$ is plotted across the phase diagrams with periodic (a) and antiperiodic (b) boundary conditions around the cylinder. We see an enhancement of the density wave component for antiperiodic boundary conditions at larger $J_2 / J_1$.

\begin{thebibliography}{99}
\bibitem{pdw_Kondo}
Short period PDW correlations with power-law fall-off were found previously in DMRG studies of asymmetric two leg ladders.\cite{fradkin2leg_1, fradkin2leg_2}
\bibitem{xi_SDW}
An alternative definition for the correlation length which we have studied is $\xi_{sdw}(\vec{Q},\vec{q}_{min}) = \frac{1}{| \vec{q}_{min} |} \sqrt{ \frac{S(\vec{Q})}{S(\vec{Q}+\vec{q}_{min})} - 1 }$ where $\vec{Q}$ is the peak and $| \vec{q}_{min} | = 2 \pi / L_{x}$.  While there are small quantitative differences that result from the use of this alternative definition, it does not make a qualitative difference.
\end{thebibliography}
\end{document}